\pdfoutput=1
\documentclass[sigconf,nonacm]{acmart}
\usepackage{xspace}
\usepackage{stmaryrd}
\allowdisplaybreaks
\usepackage{listings}

\newcommand{\relegate}{the Appendix}

\newcommand{\id}{\mathrm{id}}
\newcommand\comp/ \newcommand{\eq}{\mathit{eq}}
\newcommand{\disj}{\mathit{disj}}
\newcommand{\geqn}[2]{\mathop\geq\nolimits_{#1}#2.}
\newcommand{\leqn}[2]{\mathop\leq\nolimits_{#1}#2.}
\newcommand{\hasshape}{\mathit{hasShape}}
\newcommand{\hasvalue}{\mathit{hasValue}}
\newcommand{\test}{\mathit{test}}
\newcommand{\closed}{\mathit{closed}}
\newcommand{\lessthan}{\mathit{lessThan}}
\newcommand{\lessthaneq}{\mathit{lessThanEq}}
\newcommand{\morethan}{\mathit{moreThan}}
\newcommand{\morethaneq}{\mathit{moreThanEq}}
\newcommand{\eqlang}{\sim}
\newcommand{\uniquelang}{\mathit{uniqueLang}}
\newcommand{\iexpr}[2]{\llbracket #1 \rrbracket^{#2}}      
\newcommand{\sdef}[2]{\mathit{def}(#1,#2)}
\newcommand{\head}[1]{\mathit{head}(#1)}
\newcommand{\tail}[1]{\mathit{tail}(#1)}
\newcommand{\conc}{\cdot}
\newcommand{\graph}{\mathit{graph}}
\newcommand{\frag}{\mathit{Frag}}
\newcommand{\paths}{\mathit{paths}}
\newcommand{\pathsfrom}{\mathit{pathsfrom}}
\newcommand{\reverse}[1]{#1^-}
\newcommand{\enne}{\ \& \ }

\title{Shape Fragments}

\author{Thomas Delva}
\orcid{0000-0001-9521-2185}
\affiliation{%
  \institution{IDLab, Ghent University, imec}
  \city{Ghent}
  \country{Belgium}}
\email{thomas.delva@ugent.be}
\author{Anastasia Dimou}
\orcid{0000-0003-2138-7972}
\affiliation{%
  \institution{Dept.\ Computer Science, KU Leuven}
  \city{Leuven}
  \country{Belgium}}
\email{anastasia.dimou@kuleuven.be}
\author{Maxime Jakubowski}
\orcid{0000-0002-7420-1337}
\affiliation{%
  \institution{DSI, Hasselt University}
  \city{Hasselt}
  \country{Belgium}}
\email{maxime.jakubowski@uhasselt.be}
\author{Jan Van den Bussche}
\orcid{0000-0003-0072-3252}
\affiliation{%
  \institution{DSI, Hasselt University}
  \city{Hasselt}
  \country{Belgium}}
\email{jan.vandenbussche@uhasselt.be}

\begin{abstract}

  In constraint languages for RDF graphs, such as ShEx and SHACL,
  constraints on nodes and their properties in RDF graphs are
  known as ``shapes''.  Schemas in these languages list the
  various shapes that certain targeted nodes must satisfy for the
  graph to conform to the schema.  Using SHACL, we propose in
  this paper a novel use of shapes, by which a set of shapes is
  used to extract a subgraph from an RDF graph, the so-called
  shape fragment.  Our proposed mechanism fits in the framework
  of Linked Data Fragments.  In this paper, (i)~we define our
  extraction mechanism formally, building on recently proposed
  SHACL formalizations; (ii)~we establish correctness properties,
  which relate shape fragments to notions of provenance for
  database queries; (iii)~we compare shape fragments with SPARQL
  queries; (iv)~we discuss implementation options; and (v)~we
  present initial experiments demonstrating that shape fragments
  are a feasible new idea.

\end{abstract}

\ccsdesc[500]{Information systems~Semantic web description
languages}
\ccsdesc[500]{Information systems~Query languages}
\ccsdesc[500]{Theory of computation~Data provenance}

\keywords{data on the Web, Linked Data Fragments, SHACL,
provenance}

\begin{document}

\settopmatter{printfolios=true}
\maketitle

\section{Introduction}

This paper proposes and investigates the use of \emph{shapes} to
retrieve subgraphs from RDF graphs.  The Resource Description
Framework (RDF) \cite{rdf11primer} is a data model, often used on
the Web, to represent data as sets of subject--property--object
triples.  Viewing properties as edge labels, such data is
naturally interpreted as a labeled graph.

Shapes are constraints on nodes (IRIs or literals) and their
properties in the context of an RDF graph.  These constraints can
be complex and can traverse the graph, checking
properties of properties, and so on.  Shapes are specified in
schemas, used to validate RDF graphs; a popular language for
writing such schemas is the W3C recommended language SHACL
\cite{shacl}.\footnote{The other popular schema language for RDF
data is ShEx \cite{shex}, which we will discuss later in this
paper.} In SHACL, schemas are called ``shapes graphs'', but we
will mostly refer to them simply as \emph{shape schemas}.

Specifically, a shape schema specifies a set of shapes, each
associated with a \emph{target}, which is a simple type of
node-returning query.  Each such target--shape pair states
that all nodes returned by the target satisfy the shape.  When
these inclusions hold in an RDF graph, the graph is said to
\emph{conform} to the schema.

\begin{example} \label{intro-ex-shape}
A node on which a shape is evaluated is referred to as a \emph{focus node}.
We give two simple examples of shapes (for now just expressed in English)
about nodes in a publication graph:
\begin{description}
\item[Shape $s_1$:] ``The focus node has at least one
\textsf{author} property, and at least one \textsf{journal}
or at least one \textsf{conference} property.''
\item[Shape $s_2$:] ``The focus node has
at least one \textsf{editor} property.''
\end{description}
Targets are simple forms of queries such as:
\begin{description}
\item[Target $t_1$:] ``All nodes of type \textsf{paper}.''
\item[Target $t_2$:] ``All nodes that are
the \textsf{conference} property of some node.''
\end{description}
For example, a shape schema may specify the two
shapes $s_1$ and $s_2$, and associate target $t_1$ to $s_1$ and
$t_2$ to $s_2$.
A graph then conforms to this shape graph if every node
returned by $t_j$ satisfies shape $s_j$, for $j=1,2$.
\qed
\end{example}

Shape schemas play the same role for RDF graphs as database
schemas and integrity constraints do for databases.  By
specifying the intended structural constraints on the data, they
help maintaining data quality \cite{validatingrdf}.  The shape
schema can also be used by the query optimizer in processing
SPARQL queries \cite{SelectivityEstimation,ShapeStats}.
Knowledge of a schema helps data consumers to effectively
formulate their SPARQL queries in the first place.  Moreover, a
shape schema may be specified at the data consumer's side,
expressing the structural constraints required for the data to be
usable by local applications (e.g., \cite{TyCuS,GDPRtEXT}).

SHACL is a rather powerful language, allowing the expression of
quite complex shapes.  However, the task of checking whether an
input RDF graph conforms to a shape schema can, at least in
principle, also be done by evaluating a, typically complex,
SPARQL query.\footnote{SPARQL is the W3C recommended query
language for RDF data. Our statement is taking exception of
\emph{recursive} shape schemas \cite{andresel,bj-recshacl}, but
note that also non-recursive SHACL is already quite expressive.
Also, recursive SHACL is not yet standardized.}  That SHACL was
still invented and proposed within the Web community strongly
indicates that this language allows for easier expression of
typical constraints on the structure of the data.  This suggests
that there may be an opportunity to leverage SHACL beyond
conformance checking, and use it also to retrieve data.

In this paper, we explore this opportunity.  We certainly do not
consider SHACL to be a substitute for a general-purpose RDF query
language.  Instead, we specifically focus on
\emph{subgraph-returning} queries.  While SPARQL can define such
queries too, we propose that shapes and shape schemas can offer
an attractive alternative.

Concretely, we propose the following mechanism.
\begin{description}

  \item[Shape fragment:] Let $S$ be a set of shapes, which we may
    refer to as \emph{request shapes}.  It is natural to consider
    the query that simply returns all nodes in the input graph $G$ that
    satisfy at least one of the request shapes. However, we can
    go beyond that, and retrieve the subgraph of $G$ formed by
    tracing out, for each node $v$ satisfying some request shape,
    the \emph{neighborhood} of $v$ in $G$, as prescribed by
    that shape.  We call the subgraph thus formed the \emph{shape
    fragment} of $G$ for $S$.

  \item[Schema fragment:]
    A special case of shape fragment that
    we expect to be common is when the request shapes are taken
    from a shape schema $H$ to which $G$ is known to conform.
    Specifically, for the request shapes, we take the shapes from the
    schema, conjoined with their targets.  The resulting shape
    fragment of $G$ with respect to $H$,
    will then consist of the neighborhoods of all
    target nodes, as prescribed by the shape associated to the
    target.

\end{description}

To specify the above mechanism precisely, we need to give a
formal definition of the neighborhood of a node in a graph with
respect to some shape.  Finding the ``right'' definition
has been one of the main goals of the present work.  This is not
easy, since shapes can amount to powerful logical expressions,
involving universal quantifiers and negation.  This will be
discussed in detail later in the paper; here, we just continue
our simple example.

\begin{example}
  For the shape schema from Example~\ref{intro-ex-shape}, the
  shape fragment of a graph would consist of
  all nodes of type paper, together with their outgoing edges
  labeled \textsf{author}, \textsf{journal} or
  \textsf{conference}; these outgoing edges comprise the
  neighborhood for target--shape pair $(t_1,s_1)$.
  Furthermore, the shape fragment
  will contain all nodes that are the
  conference of some node, together with their incoming edges
  labeled \textsf{conference}, and outgoing edges labeled
  \textsf{editor}; these edges comprise the neighborhood for
  target--shape pair $(t_2,s_2)$.
  \qed
\end{example}

This paper is organized as follows.
We begin in Section~\ref{secspec} with an informal introduction to shape
fragments, and explain how they fit the framework of Linked Data
Fragments (LDF \cite{ldf,tpf,brtpf,smartkg}).
Section~\ref{secformal} gives a self-contained formal definition
of shape fragments, revolving around our definition of
neighborhood.  We work from a formalization of SHACL proposed by
Corman, Reutter and Savkovic \cite{corman}, which is gaining
traction \cite{andresel,leinberger,shaclsatsouth}. Crucially, we
prove that our definition of neighborhoods is \emph{correct} in
the following sense:
\begin{description}
  \item[Correctness for shape fragments:] We show that a node $v$
    satisfies a request shape in the shape fragment,
    if and only if $v$ satisfies that shape in the original graph.
    Thus, neighborhoods consist of exactly the right information
    to satisfy shapes.
  \item[Correctness for schema fragments:] If a graph $G$
    conforms to a schema $H$, then we show that the
    shape fragment of $G$ with respect to $H$
    still conforms to $H$, as one would expect.
\end{description}

In Section~\ref{secimpl}, we explore how shape fragments can be
implemented, either by translation to SPARQL, or by instrumenting
an existing SHACL validator.  We present initial experiments
showing that computation of shape fragments is feasible.  In
this paper we mainly introduce the idea of shape fragments; a
detailed investigation on processing strategies for shape
fragments is an obvious direction for further research.

Section~\ref{secrelated} relates shape fragments to data
provenance, and compares the expressive power of
shape fragments to Triple Pattern
Fragments \cite{tpf}.  We also compare to a recent proposal,
similar to shape fragments, made by Labra Gayo
\cite{labra-subsets}, which appeared
independently of our own work.  We believe that two independent
researchers or research groups proposing a similar idea can
underline that the idea is indeed natural.
Section~\ref{seconcl} concludes the paper by discussing
topics for further research.

\section{Shape Fragments} \label{secspec}

\begin{figure*}
    \centering
    \includegraphics[width=0.8\linewidth]{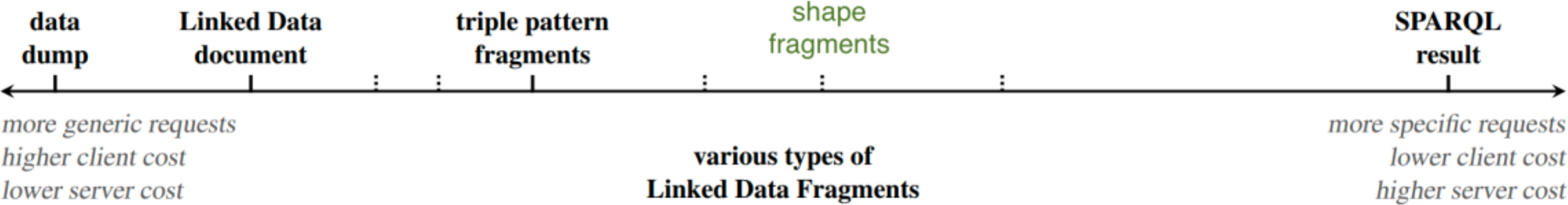}
    \caption{Positioning shape fragments in the LDF Framework
    (adapted from \cite{tpf}).}
    \label{fig:shapeFragments}
\end{figure*}

In this Section we give an introduction to shape fragments for
readers having already some familiarity with RDF and SHACL\@.  A
self-contained formal development is given in the next Section.
Moreover, we have defined a complete specification of shape fragments
which closely follows the existing W3C SHACL recommendation, and
explains in detail how each construct of core SHACL contributes
to the formation of the shape fragment \cite{attachment}.

As a first example of a shape, consider data for a student-oriented
workshop, where we require that every workshop paper has at least
one author of type student.  This constraint is expressed by the
following shape--target pair in SHACL:

\begin{verbatim}
:WorkshopShape sh:targetClass :WorkshopPaper;
  sh:property [ 
    sh:path :author; sh:qualifiedMinCount 1 ; 
    sh:qualifiedValueShape [ sh:class :Student ] ] .
\end{verbatim}

The first triple shown above specifies the target, in this case,
all nodes of type \textsf{:WorkshopPaper}.
The remaining triples specify the shape itself.
Shapes express constraints on individual nodes, called focus nodes.
In this example, the shape
expresses that the focus node should be related, through property
\textsf{:author}, to at least one node of type \textsf{:Student}.

An RDF graph conforms
to a shape--target pair if every node specified by the target
conforms to the shape.
For example, consider the following RDF graph in Turtle syntax
\cite{rdf11turtle}:
\begin{verbatim}
:p1 a :WorkshopPaper ; :author :Anne, :Bob, :Alice .
:p2 a :WorkshopPaper ; :author :Anne, :Bob .
:Anne a :Professor . :Bob a :Professor . 
:Alice a :Student .
\end{verbatim}

Paper \textsf{:p1} has author \textsf{:Alice}, who is a student,
so \textsf{:p1} conforms to the shape.  However, \textsf{:p2}
clearly violates the shape.  Hence, the graph does not conform to
the shape--target pair.

In general, a \emph{shapes graph} in SHACL is a collection of
shape--target pairs.  The task of validating an RDF graph, with
respect to a shapes graph, produces a validation report.
The report lists, for each shape--target pair, the target nodes
that violate the shape.  When the report is empty, the RDF graph
conforms to the shapes graph.

In the present paper, we refer to shapes graphs as shape
schemas, and propose that they can also be used
to \emph{retrieve} information.  Roughly speaking, we retrieve,
for each shape--target pair, all target
nodes conforming to the shape, including all triples ``tracing
out the shape''.  We call these triples the \emph{neighborhood}
of the conforming node, and we refer to the union of all
neighborhoods as the \emph{shape fragment} for the schema.

For our example \textsf{:WorkshopShape}, the neighborhood of a
conforming node $v$ would consist of all triples ($v$
\textsf{:author} $x$) from the graph where $x$ is of type
\textsf{:Student}, implying that, for each such $x$, also the
triple ($x$ \textsf{a :Student}) is included in the
neighborhood.  In SPARQL terms, the shape fragment for our
simple example schema corresponds to the result of the query
\begin{verbatim}
CONSTRUCT WHERE
{ ?v a :WorkshopPaper . ?v :author ?x . ?x a :Student }
\end{verbatim}

If we start from a data graph that conforms to the schema, the shape
fragment will also conform to this schema.
For example, consider the following shape schema:
\begin{verbatim}
:AddressShape a sh:NodeShape ;
  sh:property [
    sh:path :postalCode ;
    sh:datatype xsd:string ;
    sh:maxCount 1 ; ] .

:PersonShape a sh:NodeShape ;
  sh:targetClass :Person ;
  sh:property [
    sh:path :address ;
    sh:minCount 1 ;
    sh:node :AddressShape ;  ] .
\end{verbatim}
On a conforming data graph,
the shape fragment will consist of the neighborhoods of
all nodes of target type \textsf{:Person}
according to \textsf{:PersonShape}.
This shape expresses that the focus node must be the subject of
at least one \textsf{:address} triple; moreover, all objects of
such triples must conform to \textsf{:AddressShape}.  The latter
shape (which is only auxiliary; it has no target of its own)
expresses that the focus node can be the subject of at most
one \textsf{:postalCode} triple; moreover, if this triple is
present, the object must be a literal of type
\textsf{xsd:string}.  Thus, the shape fragment will consist of
all triples of the form ($v$ \textsf{a :Person}) in the graph,
plus, for each such $v$,
all triples ($v$ \textsf{:address} $x$) in the graph, plus, for
each such $x$, a triple ($x$ \textsf{:postalCode} $c$) if present
in the graph.

SHACL is a rather powerful language; shapes can be negated, can
test for disjointness, etc.  For example, consider:
\begin{verbatim}
:HappyAtWork a sh:NodeShape ;
  sh:not [
    sh:path :friend ;
    sh:disjoint :colleague ; ] .
\end{verbatim}
This shape expresses that the focus node has at least one friend
who is also a colleague.  The shape fragment would retrieve, for
every conforming node $v$, the union of all pairs of triples ($v$
\textsf{:friend} $x$) and ($v$ \textsf{:colleague} $x$) for each
common $x$ that exists in the data graph.

Shape fragments fit the framework of Linked Data Fragments
\cite{ldf,tpf,brtpf,smartkg} (LDF)\@ for publication interfaces
to retrieve RDF (sub)graphs.  At one end of the spectrum the
complete RDF graph is retrieved; at the other end, the results of
arbitrary SPARQL queries.  Triple Pattern Fragments (TPF)
\cite{tpf}, for example, represent an intermediate point where
all triples from the graph that match a given SPARQL triple
pattern are returned.

On this spectrum, shape fragments lie between TPF and arbitrary
SPARQL, as shown in Fig.~\ref{fig:shapeFragments}, taking
advantage of the merits of both approaches.  On the one hand,
shape fragments may reduce the server cost, similarly to TPF, but
they can also perform fewer requests as multiple TPFs can be
expressed as a single shape fragment.  On the other hand, shape
fragments may also perform quite powerful requests, similarly to
SPARQL endpoints, but without reaching the full expressivity of
SPARQL.  In fact, as opposed to TPF and SPARQL endpoints that
predetermine where the cost will be, our proposed shape fragments
mechanism may be implemented both on client- or server-side,
adjusting and balancing the costs.

By returning a subgraph rather than a set of
variable bindings (as in SPARQL select-queries),
results can be better compressed \cite{smartkg}.
Returning subgraphs also simplifies keeping track of provenance
\cite{glavic-data-provenance} or footprint \cite{rubenshapeblog}.
Trading off efficiency against expressive power, processing shape fragments
may be more efficient than processing arbitrary SPARQL
(and with fewer requests than in the case of TPF)
although this needs to be confirmed by further research.
A final advantage of shape fragments is software reuse;
shapes now serve the double purpose of describing the available data,
and of accessing the data.


\section{Formal definition and sufficiency} \label{secformal}

In this section, we first 
present a formalization of SHACL, following
the work by Corman, Reutter and Savkovic~\cite{corman}.
We extend their formalization to cover all
features of SHACL, such as disjointness, zero-or-one
property paths, closedness, language tags, node tests, and
literals.  We then proceed to the formal definition of
neighborhoods and shape fragments, and establish an important
correctness result, called the \emph{sufficiency
property.}

\theoremstyle{acmdefinition}
\newtheorem{remark}[theorem]{Remark}

\subsection{SHACL formalization} \label{secsh:formal}

We assume three pairwise disjoint infinite sets $I$, $L$, and $B$
of \emph{IRIs}, \emph{literals}, and \emph{blank nodes},
respectively.  We use $N$ to denote the union $I \cup B \cup L$;
all elements of $N$ are referred to as \emph{nodes}.
Literals may have a ``language tag'' \cite{rdf11primer}.  We abstract this by
assuming an equivalence relation $\eqlang$ on $L$, where
$l \eqlang l'$ represents that $l$ and $l'$ have
the same language tag.  Moreover, we assume a
strict partial order $<$ on $L$ that abstracts comparisons between
numeric values, strings, dateTime values, etc.

An \emph{RDF triple} $(s, p, o)$ is an element of $(I \cup B)
\times I \times N$. We refer to the elements of the triple as the
subject $s$, the property $p$, and the object $o$.  An \emph{RDF
graph} $G$ is a finite set of RDF triples.  It is natural to
think of an RDF graph as an edge-labeled, directed graph, viewing
a triple $(s,p,o)$ as a $p$-labeled edge from node $s$ to node $o$.

We formalize SHACL property paths as \emph{path expressions} $E$.
Their syntax is given by the following grammar, where $p$ ranges
over $I$:
\[
  E ::= p \mid E^- \mid E_1 \comp E_2 \mid E_1 \cup E_2 \mid E^* \mid E?
\]

SHACL can do many tests on individual nodes, such as testing whether
a node is a literal, or testing whether an IRI matches some
regular expression.  We abstract this by assuming a set
$\Omega$ of \emph{node tests}; for any node test $t$ and node
$a$, we assume it is well-defined whether or not $a$
\emph{satisfies} $t$.

The formal
syntax of \emph{shapes} $\phi$ is now given by the following grammar.
\begin{tabbing}
  $F ::= E \mid \id$ \\
$\phi$ \=$::=$ \=$\,\top \mid \bot \mid \hasshape(s) \mid \test(t) \mid
\hasvalue(c)$ \\
\>${}\mid{}$\>$\eq(F, p) \mid \disj(F, p) \mid \closed(P)$\\
\>${}\mid{}$\>$\lessthan(E, p) \mid \lessthaneq(E, p) \mid
  \uniquelang(E)$\\
\>${}\mid{}$\>$\neg \phi \mid \phi \land \phi \mid \phi \lor \phi$\\
\>${}\mid{}$\>$\geqn{n}{E}{\phi} \mid \leqn{n}{E}{\phi} \mid
\forall E.\phi$
\end{tabbing}
with
$E$ a path expression;
$s \in I \cup B$;
$t \in \Omega$;
$c \in N$;
$p \in I$;
$P \subseteq I$ finite;
and
$n$ a natural number.

\begin{remark}
  Note that in shapes of the form $\eq(F,p)$ or $\disj(F,p)$, the
  argument expression $F$ can be either a path expression $E$ or
  the keyword `id'.  We will see soon that `id' stands for the
  focus node.  We need to include these id-variants
in our formalisation, to reflect
  the distinction made in the
  SHACL recommendation between ``node shapes'' (expressing
  constraints on the focus node itself) and
``property shapes'' (expressing constraints on nodes reachable
  from the focus node by a path expression). \qed
\end{remark}

We formalize SHACL shapes graphs as \emph{schemas}.  We first
define the notion of \emph{shape definition}, as a triple $(s,
\phi, \tau)$ where $s \in I \cup B$, and $\phi$ and $\tau$ are
shapes. The elements of the triple are referred to as the
\emph{shape name}, the \emph{shape expression}, and the
\emph{target expression}, respectively.\footnote{Real SHACL only
supports specific shapes for targets, but our development works
equally well when allowing any shape for a target.}

Now a \emph{schema} is a finite set $H$ of shape definitions such
that no two shape definitions have the same shape name.
Moreover, as in the current SHACL recommendation, in this paper
we consider only \emph{nonrecursive} schemas.  Here,
$H$ is said to be recursive if there is a directed cycle in the
directed graph formed by the shape names, with an edge $s_1 \to
s_2$ if $\hasshape(s_2)$ occurs in the shape expression defining
$s_1$.

In order to define the semantics of shapes and shape schemas, we
first recall that a path expression $E$ evaluates on an RDF graph
$G$ to a binary relation on $N$, denoted by $\iexpr
EG$ and defined as follows.
$\iexpr{p}{G} = \{(a,b) \mid (a,p,b) \in G\}$; 
$\iexpr{E^-}{G} = \{(b,a) \mid (a,b) \in \iexpr{E}{G}\}$;
$\iexpr{E?}{G} = \{(a,a) \mid a \in N\} \cup \iexpr{E}{G}$;
$\iexpr{E_1 \cup E_2}{G} = \iexpr{E_1}{G} \cup \iexpr{E_2}{G}$;
$\iexpr{E_1 \comp E_2}{G} = \{(a,c) \mid \exists b : (a,b) \in
\iexpr{E_1}{G}$ \& $(b,c) \in \iexpr{E_2}{G}\}$; and
$\iexpr{E^*}{G} = \text{the}$ reflexive-transitive closure of
$\iexpr{E}{G}$.  Finally, we also define $\iexpr{\id}G$, for any
$G$, to be simply the identity relation on $N$.

\begin{table}
\caption{Conditions for conformance of a node to a shape.}
\label{tabdef}
\centering
\begin{tabular}{ll}
\toprule
$\phi$ & $H,G,a \models \phi$ if: \\
\midrule
$\hasvalue(c)$ & $a=c$ \\
$\test(t)$ & $a$ satisfies $t$ \\
$\hasshape(s)$ & $H,G,a \models \sdef sH$ \\
$\geqn{n}{E}{\psi}$ &
$\sharp \{b \in \iexpr{E}{G}(a) \mid H,G,b \models \psi\} \geq n$ \\
$\leqn{n}{E}{\psi}$ &
$\sharp \{b \in \iexpr{E}{G}(a) \mid H,G,b \models \psi\} \leq n$ \\
$\forall E.\psi$ & every $b \in
\iexpr{E}{G}(a)$ satisfies $H,G,b\models \psi$ \\
$\eq(F,p)$ &
the sets $\iexpr{F}{G}(a)$ and $\iexpr{p}{G}(a)$ are equal \\
$\disj(F,p)$ &
the sets $\iexpr{F}{G}(a)$ and $\iexpr{p}{G}(a)$ are disjoint \\
$\closed(P)$ & for all triples $(a,p,b) \in G$
  we have $p \in P$ \\
$\lessthan(E,p)$ &
$b<c$ for all $b \in \iexpr{E}{G}(a)$ and $c \in \iexpr{p}{G}(a)$ \\
$\lessthaneq(E,p)$ &
$b\leq c$ for all $b \in \iexpr{E}{G}(a)$ and $c \in \iexpr{p}{G}(a)$ \\
$\uniquelang(E)$ & $b \nsim c$ for all $b\neq c \in \iexpr{E}{G}(a)$. \\
\bottomrule
\end{tabular}
\end{table}

We are now ready to define when a focus node $a$ \emph{conforms} to a
shape $\phi$ in a graph $G$, in the context of a schema $H$,
denoted by $H,G,a \models \phi$.  For the boolean operators
$\top$ (true), $\bot$ (false), $\neg$ (negation), $\land$
(conjunction), $\lor$ (disjunction), the definition is obvious.
For the other constructs, the definition is shown in
Table~\ref{tabdef}.  In this table, we employ the following
notations:
\begin{itemize}
  \item
In the definition for $\hasshape(s)$ we use
the notation $\sdef sH$ to denote the shape expression defining
shape name $s$ in $H$.  If $s$ does not have a definition in $H$,
we let $\sdef sH$ be $\top$ (this is the behavior in real
SHACL)\@.
\item
    We use the notation $R(x)$, for a binary relation
$R$, to denote the set $\{y \mid (x,y) \in R\}$. We apply this
notation to the case where $R$ is of the form $\iexpr EG$ and $x$
    is a node.  For example, $\iexpr{\id}G(a)$ equals the
    singleton $\{a\}$.
  \item
    We also use the notion $\sharp X$ for the cardinality
of a set $X$.
\end{itemize}
    Note that the conditions for
$\lessthan(E,p)$ and $\lessthaneq(E,p)$ imply that $b$ and $c$
must be literals.

In \relegate,
we show that our formalization fully covers real SHACL\@.

\begin{example}
  \begin{itemize}
\item
  The shape \textsf{:WorkshopShape} from Section~2 can be
  expressed in the formalisation as follows:
  $$ \geqn 1{\textsf{{:}author}}{\geqn
  1{\mathsf{rdf{:}type}/\mathsf{rdf{:}subclassOf}^*}
  {\hasvalue(\textsf{{:}Student})}}
  $$
\item
  The shape \textsf{:HappyAtWork} from Section~2 is
  expressed as $\neg
  \disj(\textsf{{:}friend},\textsf{{:}colleague})$.
\item
  For an IRI $p$, the shape $\neg \disj(\id,p)$ expresses
  that the focus node has a $p$-labeled self-loop, and the shape
  $\eq(\id,p)$ expresses that its \emph{only} outgoing $p$-edge
  is a self-loop.
  \qed
\end{itemize}
\end{example}

Finally, we can define conformance of a graph to a schema as
follows.  RDF graph $G$ \emph{conforms} to schema $H$ if for
every shape definition $(s, \phi, \tau) \in H$ and for every $a
\in N$ such that $H,G,a \models \tau$, we have $H,G,a \models
\phi$.

\begin{remark}
Curiously, SHACL provides shapes $\lessthan$ and $\lessthaneq$
  but not their variants $\morethan$ and $\morethaneq$ (with the
  obvious meaning).  Note
  that $\morethan(E,p)$ is not equivalent to $\neg
  \lessthaneq(E,p)$.  In this paper we stay with the SHACL
  standard, but our treatment is easily extended to
$\morethan$ and $\morethaneq$.
\end{remark}

\subsection{Neighborhoods} \label{secneigh}

\begin{table*}
\caption{Neighborhood in the context of a schema $H$, when $G,v
\models \phi$ and $\phi$ is in negation normal form.
  In particular, in rules 2 and 6, we assume
  that $\neg\sdef sH$ and $\neg\psi$ are put in negation normal form.
In the omitted cases, and when $G,v \not\models \phi$,
  the neighborhood is defined to be empty.}
\label{tabdefrag}
\centering
\begin{tabular}{ll}
\toprule
$\phi$ & $B(v,G,\phi)$ \\
\midrule
$\hasshape(s)$ & $B(v,G,\sdef sH)$ \\
$\neg \hasshape(s)$ & $B(v,G,\neg \sdef sH)$ \\
$\phi_1 \land \phi_2$ & $B(v,G,\phi_1) \cup B(v,G,\phi_2)$ \\
$\phi_1 \lor \phi_2$ & $B(v,G,\phi_1) \cup B(v,G,\phi_2)$ \\
$\geqn{n}{E}{\psi}$ & $\bigcup \{ \graph(\paths(E,G,v,x)) \cup
B(x,G,\psi) \mid (v,x) \in \iexpr EG \ \& \ G,x \models \psi\}$ \\
$\leqn{n}{E}{\psi}$ & $\bigcup \{ \graph(\paths(E,G,v,x)) \cup
B(x,G,\neg\psi) \mid (v,x) \in \iexpr EG \ \& \ G,x \models \neg\psi\}$ \\
$\forall E.\psi$ & $\bigcup \{ \graph(\paths(E,G,v,x)) \cup
B(x,G,\psi) \mid (v,x) \in \iexpr EG\}$ \\
$\eq(E,p)$ & $\graph(\pathsfrom(E\cup p,G,v))$ \\
$\eq(\id,p)$ & $\{(v,p,v)\}$ \\
$\neg\eq(E,p)$ & $\graph(\{\pi \in \pathsfrom(E,G,v) \mid
(v,p,\head\pi) \notin G\}) \cup \{(v,p,x) \in G \mid (v,x) \notin
\iexpr EG\}$ \\
$\neg \eq(\id,p)$ & $\{(v,p,x) \in G \mid x \neq v\}$ \\
$\neg\disj(E,p)$ & 
$\bigcup\{\graph(\paths(E,G,v,x)) \cup \{(v,p,x)\} \mid (v,x) \in
\iexpr EG \enne (v,p,x)\in G\}$ \\
$\neg \disj(\id,p)$ & $\{(v,p,v)\}$ \\
$\neg \lessthan(E,p)$ &
$\bigcup\{\graph(\paths(E,G,v,x)) \cup \{(v,p,y)\} \mid (v,x) \in
\iexpr EG \enne (v,p,y)\in G \enne x \nless y\}$ \\
$\neg \lessthaneq(E,p)$ &
$\bigcup\{\graph(\paths(E,G,v,x)) \cup \{(v,p,y)\} \mid (v,x) \in
\iexpr EG \enne (v,p,y)\in G \enne x \nleq y\}$ \\
$\neg\uniquelang(E)$ & $\graph(\{\pi \in \pathsfrom(E,G,v) \mid
\exists x \in \iexpr EG(v) : x \neq \head\pi \enne x \eqlang \head\pi\})$ \\
$\neg \closed(P)$ & $\{(v,p,x) \in G \mid p \notin P\}$ \\
\bottomrule
\end{tabular}
\end{table*}

The fundamental notion to be defined is that of the
\emph{neighborhood} of a node $v$ for a shape $\phi$ in a graph
$G$.  The intuition is that
this neighborhood consists of those triples in $G$ that show that
$v$ conforms to $\phi$; if $v$ does not conform to $\phi$, the
neighborhood is set to be empty.  We want a generic, tractable,
deterministic definition that formalizes this intuition.  Our
definition should also omit unnecessary triples; for otherwise,
one could simply define the neighborhood to be $G$ itself! 

Before developing the definition formally, we discuss the salient features
of our approach.
\begin{description}

\item[Negation] Following the work by Gr\"adel and Tannen on
  supporting where-provenance in the presence of negation
    \cite{gt_negation}, we assume shapes are in \emph{negation
    normal form}, i.e., negation is only applied to atomic
    shapes.  This is no restriction, since every shape can be put
    in negation normal form, preserving the overall syntactic
    structure, simply by pushing negations down.  We push
    negation through conjunction and disjunction using De
    Morgan's laws.  We push negation through quantifiers as
    follows: $$ \neg \geqn {n+1}E\psi  \equiv \leqn nE\psi \quad
    \neg \leqn nE\psi  \equiv \geqn {n+1}E\psi \quad \neg
    \forall E.\psi  \equiv \geqn 1E{\neg\psi} $$  The negation of
    $\geqn 0E\psi$ is simply false.

\item[Node tests] We leave the neighborhood for $\hasvalue$ and
  $\test$ shapes empty, as these involve no properties, i.e., no
    triples.

\item[Closedness] We also define the neighborhood for
  $\closed(P)$ to be empty, as this is a minimal subgraph in
    which the shape is indeed satisfied.  A reasonable
    alternative approach would be to return all properties of the
    node, as ``evidence'' that these indeed involve only IRIs in
    $P$.  Indeed, we will show in Section~\ref{secsuff} that our
    definitions, while minimalistic, are taken such that they can
    be relaxed without sacrificing the sufficiency property.

\item[Disjointness]  Still according to our minimal approach, the
  neighborhood for disjointness shapes is empty.  Analogously,
    the same holds for $\lessthan$ and $\uniquelang$ shapes.

\item[Equality]  The neighborhood for a shape $\eq(E,p)$ consists
  of the subgraph traced out by the $E$-paths and $p$-properties of the
    node under consideration, evidencing that the sets of
    end-nodes are indeed equal.  Here, we can no longer afford to
    return the empty neighborhood, although equality would hold
    trivially there.  Indeed, this would destroy the relaxation
    property promised above.  For example, relaxing by adversely
    adding just one $E$-path and one $p$-property with distinct
    end-nodes, would no longer satisfy equality.

\item[Nonclosure]  The neighborhood for a shape $\neg \closed(P)$
  consists of those triples from the node under consideration
    that involve properties outside $P$, as expected.

\item[Nonequality]  For $\neg \eq(E,p)$ we return the subgraph
traced out by
the $E$-paths from the node $v$ under consideration
    that end in a node that is \emph{not} a $p$-property of $v$,
    and vice versa.  A similar approach is taken for
    nondisjointness and negated $\lessthan$ shapes.

\item[Quantifiers] The neighborhood for $\forall E.\psi$
  consists, as expected, of the subgraph traced out by
  all $E$-paths from
    the node under consideration to nodes $x$, plus the
    $\psi$-neighborhoods of these nodes $x$.  For $\geqn nE\psi$
    we do something similar, but we take only those $x$ that conform
    to $\psi$.  Given the semantics of the $\mathord \geq_n$
    quantifier, it seems tempting to instead just take a
    selection of $n$ of such nodes $x$.  However, we want a
    deterministic definition of neighborhood, so we take all $x$.
    Dually, for $\leqn nE\psi$, we return the subgraph traced out
    by $E$-paths from the current node to
    nodes \emph{not} conforming to $\psi$, plus their
    $\neg\psi$-neighborhoods.
\end{description}

Towards a formalization of the above ideas, we first make precise
the intuitive notion of a path in an RDF graph, and of the
subgraph traced out by a path.  Paths are finite sequences
of adjacent steps. Each step either moves forward from the
subject to the object of a triple, or moves backward from the
object to the subject.  We make backward steps precise by
introducing, for each property $p\in I$, its \emph{reverse},
denoted by $p^-$.  The set of reverse IRIs is denoted by $I^-$.
We assume $I$ and $I^-$ are disjoint, and moreover, we also
define $(p^-)^-$ to be $p$ for every $p\in I$.

For any RDF triple $t=(s,p,o)$, the triple $t^- := (o,p^-,s)$ is called a
\emph{reverse triple}.  As for IRIs, we define $(t^-)^-$ to be $t$.
A \emph{step} is an RDF triple (a forward step) or a
reverse triple (a backward step).
For any step $t=(x,r,y)$, we refer to $x$ as the
\emph{tail}, denoted by $\tail t$, and to $y$ as the \emph{head},
denoted by $\head t$.  A \emph{path} is a nonempty finite sequence $\pi$
of steps so that $\head{t_1}=\tail{t_2}$ for any two subsequent
steps $t_1$ and $t_2$ in $\pi$.  The \emph{tail} of $\pi$ is
the tail of its first step; the \emph{head} of $\pi$ is the head
of its last step.  Any two paths $\pi$ and $\pi'$ where
$\head\pi=\tail{\pi'}$ can be concatenated; we denote this by
$\pi \conc \pi'$.

The \emph{graph} traced out by a path $\pi$, denoted by
$\graph(\pi)$, is simply the set of RDF triples underlying the
steps of the path.  Thus, backward steps must be reversed.
Formally, $$ \graph(\pi) = \{t \mid \text{$t$ forward step in
$\pi$}\} \cup \{t^- \mid \text{$t$ backward step in $\pi$}\}. $$
For a set $\Pi$ of paths, we define $\graph(\Pi) = \bigcup
\{\graph(\pi) \mid \pi \in \Pi\}$.

We are not interested in arbitrary sets of paths, but in the
set of paths generated by a path expression $E$ in an RDF graph
$G$, denoted by $\paths(E,G)$ and defined in a standard manner as
follows.
$\paths(p,G) = \{(a,r,b) \in G \mid r=p\}$;
$\paths(E/E',G) = \{\pi \conc \pi' \mid \pi \in \paths(E,G)$ \&
$\pi' \in \paths(E',G)$ \& $\tail\pi=\head{\pi'}\}$;
$\paths(E \cup E',G) = \paths(E,G) \cup \paths(E',G)$;
$\paths(E?,G) = \paths(E,G)$;
$\paths(E^*,G) = \bigcup\nolimits_{i=1}^\infty \paths(E^i,G)$;
and
$\paths(E^-,G) = \{\reverse\pi \mid \pi \in \paths(E,G)\}$. Here,
$E^i$ abbreviates $E/\cdots/E$ ($i$ times), and
$\reverse\pi = t_l^-, \dots, t_1^-$ for $\pi=t_1,\dots, t_l$.
Note that $\paths(p,G)$ is a set of length-one paths.

In order to link $E$-paths to the evaluation of shapes below, we
introduce some more notation, for any two nodes $a$ and $b$:
\setlength{\jot}{0pt}
\begin{align*}
\pathsfrom(E,G,a) & := \{\pi \in \paths(E,G) \mid \tail\pi=a\} \\
\paths(E,G,a,b) &:= \{\pi \in \pathsfrom(E,G,a) \mid 
\head\pi = b\}
\end{align*}

Note that $\graph(\pi)$, for every
$\pi \in \paths(E,G)$, is a subgraph of $G$.
This will ensure that neighborhoods and
shape fragments are always subgraphs of the original graph.
Moreover, the following observation ensures that
path expressions will have the same semantics in the fragment as
in the original graph:
\begin{proposition} \label{prop:graphpaths}
Let $F=\graph(\paths(E,G,a,b))$. Then $(a,b) \in \iexpr EG$ if
and only if $(a,b) \in \iexpr EF$.
\end{proposition}
Note that $\paths(E,G)$ may be infinite, due to the use of Kleene
star in $E$ and cycles in $G$.  However
$\graph(\paths(E,G))$ is always finite,
because $G$ is finite.

We are now ready to define neighborhoods in the context of an
arbitrary but fixed schema $H$.  To avoid clutter we will omit
$H$ from the notation.  Let $v$ be a node, $G$ be a graph, and
$\phi$ be a shape.  We define the \emph{$\phi$-neighborhood of
$v$ in $G$}, denoted by $B(v,G,\phi)$, as the empty RDF graph
whenever $v$ does not conform to $\phi$ in $G$.  When $v$ does
conform, the definition is given in Table~\ref{tabdefrag}.  As
already discussed above, by pushing negations down, we can and do
assume that $\phi$ is put in \emph{negation normal form}, meaning
that negation is only applied to atomic shapes.  (Atomic shapes
are those from the first three lines in the production
for $\phi$, in the grammar for shapes given in
Section~\ref{secsh:formal}.)

\subsection{Shape fragments and sufficiency} \label{secsuff}

The \emph{shape fragment} of an RDF graph $G$, for a finite set
$S$ of shapes, is the subgraph of $G$
formed by the neighborhoods of all nodes in $G$
for the shapes in $S$.  Formally:
$$ \frag(G,S) = \bigcup \{B(v,G,\phi) \mid v \in N \enne \phi \in
S\}. $$
Here, $v$ ranges over the universe $N$ of all nodes, but since
neighborhoods are always subgraphs of $G$, it is equivalent to
let $v$ range over all subjects and objects of triples in $G$.

The shapes in $S$ can be interpreted as arbitrary ``request
shapes''.  An interesting special case, however, is when $S$ is
derived from a shape schema $H$.  Formally, we define the shape
fragment of $G$ for $H$ as $\frag(G,H) := \frag(G,S)$, where $ S
= \{\phi \land \tau \mid \exists s : (s,\phi,\tau) \in H\}$.
Thus, the shape fragment for a schema requests the conjunction of
each shape in the schema with its associated target.

In order to state our main correctness result, we need to revisit
the definition of schema.  Recall that a schema is a set of shape
definitions, where a shape definition is of the form
$(s,\phi,\tau)$.  Until now, we allowed both the shape expression
$\phi$ and the target $\tau$ to be arbitrary shapes.  In real
SHACL, however, only shapes of the following specific forms can
be used as targets:
\begin{itemize}
  \item $\hasvalue(c)$ (node targets);
\item
    $\geqn 1{p/r^*}\hasvalue(c)$ (class-based targets: $p$ and
    $r$ stand for \textsf{type} and \textsf{subclass} from the
    RDF Schema vocabulary \cite{rdf11primer}, and $c$ is
    the class name);
  \item
    $\geqn 1p\top$ (subjects-of targets); and
\item
    $\geqn 1{p^-}\top$ (objects-of targets).
\end{itemize}
    For our purposes, however, what counts is that real
SHACL targets $\tau$ are \emph{monotone}, in the sense that if
$G,v \models \tau$ and $G \subseteq G'$, then also $G',v \models
\tau$.

We establish:
\begin{theorem}[Conformance] \label{theorem}
Assume schema $H$ has monotone targets, and assume RDF graph
$G$ conforms to $H$.  Then $\frag(G,H)$ also conforms to $H$.
\end{theorem}

In \relegate,
we prove the Conformance Theorem using the following result:
\begin{lemma}[Sufficiency] \label{lemma}
If $G,v \models \phi$ then also $G',v \models \phi$ for any RDF graph
$G'$ such that $B(v,G,\phi) \subseteq G' \subseteq G$.
\end{lemma}
We call this the Sufficiency Lemma, because of the following
corollary:
\begin{corollary} \label{corol}
Let $G$ be an RDF graph, let $S$ be a finite set of shapes, let
$\phi$ be a shape in $S$, and let $v$ be a node.  If $G,v \models
\phi$, then also $\frag(G,S),v \models \phi$.
\end{corollary}
The above corollary shows that the shape fragment $\frag(G,S)$ is
sufficient in the sense of providing provenance for any shape in
$S$ evaluated in $G$.
Thinking of a shape as a unary query, returning all nodes
that conform to it, this is exactly the ``sufficiency property'' that
has been articulated in the theory of data provenance
\cite{glavic-data-provenance}.

The Sufficiency Lemma is stated not just for the neighborhood,
but more strongly for all subgraphs that encompass the
neighborhood.  This formulation serves both a technical and a
conceptual purpose.  The technical purpose is that it helps to
prove the lemma by induction: in order to deal with universal and
maxcount quantifiers ($\forall$ and $\mathord \leq_n$), a
stronger hypothesis is needed to carry the induction through.

Moreover, the conceptual contribution of the Sufficiency Lemma is
that our definition of neighborhood is not ``take it or
leave it''.  Indeed, we allow for a shape fragment
processor to return \emph{larger} neighborhoods than the ones we
strictly define.  Theorem~\ref{theorem} and
Corollary~\ref{corol} will continue to hold.
As discussed in Section~\ref{secneigh}, this
``relaxation property'' allows the server to provide more
information, and justifies our minimalistic approach.

\begin{example} \label{exleq}
\newcommand{\fon}{\sf}
Recall the example about the student-oriented workshop from
Section~2.  As a variation, suppose each paper must have
at least one author, but can have at most one author who is
\emph{not} of type student.
These two constraints are captured by a schema $H$
with two shape definitions. One has the shape expression $\fon
\geqn 1{author}\top$, and the other has the shape expression
$$ \fon \leqn 1{author}\neg\geqn 1{type}student, $$ which in
negation normal form becomes $\fon \leqn 1{author}\leqn
0{type}student$.  Both shape definitions have target $\fon \geqn
1{type}paper$.  We denote the two shape expressions by $\phi_1$
and $\phi_2$, and the target by $\tau$.

\newcommand{\pff}{\textsf{p1}}
Consider the simple graph $G$ consisting of a single paper, say \pff.
This paper has two authors: Anne, who is a professor, and Bob,
who is a student.  Formally, $G$ consists of the five triples
$\fon (\pff,type,paper)$, $\fon (\pff,auth,Anne)$, $\fon
(\pff,auth,Bob)$, $\fon (Anne,type,prof)$ and
$\fon (Bob,type,\allowbreak student)$.

The neighborhood of \pff\ for $\phi_1 \land \tau$ consists of the
three triples
$\fon (\pff,type,paper)$, $\fon (\pff,auth,Anne)$ and $\fon
(\pff,auth,Bob)$.
The neighborhood of \pff\ for $\phi_2 \land \tau$ consists of the
three triples 
$\fon (\pff,type,paper)$, $\fon (\pff,auth,Bob)$
and $\fon (Bob,type,student)$.
Thus, $\frag(G,H)$, being the union of the two neighborhoods, 
consists of the four triples 
$\fon (\pff,type,paper)$, $\fon (\pff,auth,Anne)$, $\fon
(\pff,auth,Bob)$ and $\fon (Bob,type,\allowbreak student)$.

Note that the triple
$\fon (Bob,type,\allowbreak
student)$ is essential in the neighborhood for $\phi_2 \land
\tau$; omitting it from the shape fragment would break
conformance to $H$.
On the other hand, we are free to add the triple
$\fon (Anne,type,prof)$ to the fragment without breaking
conformance.

Finally, note that we could add to $G$ various other triples
unrelated to the shapes in $H$.  The shape fragment would omit
all this information, as desired.
\qed
\end{example}

\begin{example}
For monotone shapes, the converse of Corollary~\ref{corol}
clearly holds as well.  In general, however, the converse does
not always hold.  For example, consider the shape $\phi = \leqn
0p\top$ (``the node has no property $p$''),
and the graph $G = \{(a,p,b)\}$.  Then
the fragment
$\frag(G,\{\phi\})$ is empty, so $a$ trivially conforms to $\phi$
in the fragment.  However, $a$ clearly does not conform to $\phi$ in $G$.
\end{example}

\section{Implementation and experimental validation} \label{secimpl}

In this section we show that shape fragments can be effectively
computed, and report on initial experiments.

\subsection{Translation to SPARQL} \label{secsparql}

Our first approach to computing shape fragments is by translation
into SPARQL, the recommended query language for RDF graphs
\cite{sparql1.1}.  SPARQL select-queries return sets of
\emph{solution mappings}, which are maps $\mu$ from finite sets of
variables to $N$.  Variables are marked using question marks.
Different mappings in the result may have different domains
\cite{semanticsparql,perez_sparql_tods}.

Shape fragments are unions of neighborhoods, and neighborhoods
in an RDF graph $G$ are unions of subgraphs of the form
$\graph(\paths(E,G,a,b))$, for path expressions $E$ mentioned in
the shapes, and selected nodes $a$ and $b$.  Hence, the following
lemma is important.  For any RDF graph $G$, we denote by $N(G)$
the set of all subjects and objects of triples in $G$.
\begin{lemma} \label{lemE}
For every path expression $E$, there exists a SPARQL select-query
$Q_E(?t,?s,?p,?o,?h)$ such that for every RDF graph $G$:
\begin{enumerate}
\item
The binary relation
$\{(\mu(?t),\mu(?h)) \mid \mu \in Q_E(G)\}$ equals
$\iexpr EG$, restricted to $N(G)$.
\item
For all $a,b \in N(G)$, the RDF graph
\begin{multline*}
\{(\mu(?s),\mu(?p),\mu(?o)) \mid \mu \in Q_E(G) \enne
(\mu(?t),\mu(?h))=(a,b) \\
\& \ \text{$\mu$ is defined on $?s$, $?p$ and $?o$}\}
\end{multline*}
equals $\graph(\paths(E,G,a,b))$.
\end{enumerate}
\end{lemma}

We emphasize that the above Lemma is not obvious.  While SPARQL
queries, through property paths, can readily test if $(a,b) \in
\iexpr EG$, it is not obvious one can actually return
$\graph(\paths(E,G,a,b))$.  The detailed proof is in \relegate;
the following example gives an idea of the
proof on a simpler case.

\begin{example} \label{ex-qrpaths}
For IRIs $a$, $b$, $q$ and $r$, the following SPARQL query,
  applied to any graph $G$, returns $\graph(\paths((q/r)^*,G,a,b))$:
{\sf
\begin{tabbing}
\qquad\=\kill
SELECT ?s ?p ?o \\
WHERE \{ $a$ $(q/r)$* ?t . ?h $(q/r)$* $b$ . \{\+\\
\{ \=SELECT ?t (?t AS ?s) ($q$ AS ?p) ?o ?h \\
   \>WHERE \{ ?t  $q$ ?o . ?o $r$ ?h \} \} \\
UNION\\      
\{ SELECT ?t ?s ($r$ AS ?p) (?h AS ?o) ?h \\
 \>WHERE \{ ?t  $q$ ?s . ?s $r$ ?h \} \} \} \}
\end{tabbing}}
\qed
\end{example}

Using Lemma~\ref{lemE}, and expressing the definitions from
Table~\ref{tabdefrag} in SPARQL, we obtain a further result as
  follows. (Proof in \relegate.)
\begin{lemma} \label{lemB}
For every shape $\phi$, there exists a SPARQL select-query
$Q_\phi(?v,?s,\allowbreak ?p,\allowbreak ?o)$
such that for every RDF graph $G$,
\begin{multline*}
\{(\mu(?v),\mu(?s),\mu(?p),\mu(?o)) \mid \mu \in Q_\phi(G)\} \\
{} = \{(v,s,p,o) \in N^4 \mid (s,p,o) \in B(v,G,\phi)\}
\end{multline*}
\end{lemma}

\begin{remark}
The above Lemma should not be confused with the known result
  \cite[Proposition 3]{shaclsparql} that SPARQL can compute the set of nodes
  that \emph{conform} to a given shape.  Our result states that
  also the neighborhoods can be computed.
\end{remark}

The above two Lemmas, combined with the definition of shape
fragments, now imply the announced result:
\begin{proposition} \label{propshaclsparql}
For every finite set $S$ of shapes, there exists a SPARQL
select-query $Q_S(?s,?p,?o)$ such that for every RDF graph $G$,
$$
\{(\mu(?s),\mu(?p),\mu(?o)) \mid \mu \in Q_S(G)\} =
\frag(G,S). $$
\end{proposition}

Query expressions for shapes can quickly become quite complex,
even for just retrieving the nodes that satisfy a shape. For the
simple example shape \textsf{:PersonShape} from
Section~\ref{secspec}, such a query needs to retrieve persons
with at least one address, which is just a semijoin, but must
also test that all addresses have at most one postal code, which
requires at least a not-exists subquery involving a non-equality
join.  Shapes involving equality constraints require nested
not-exists subqueries in SPARQL, and would benefit from specific
operators for set joins, e.g.,
\cite{moerkotte_setjoin,mamoulis_setjoin}.  Shapes of the form
$\leqn 5 p \top$ requires grouping the $p$-properties and applying
a condition $\mathrm{count} \leq 5$, plus a union
with an outer join to retrieve the nodes without any
$p$-property.  Such shapes would benefit from specific operators
for group join \cite{groupjoin1,groupjoin2}.

Obviously, queries that actually retrieve the shape fragment are
no simpler.  Our Proposition~\ref{propshaclsparql} only states
that SPARQL is sufficient in principle, and leaves query
optimization for future work.

\begin{example}
  For IRIs $p$, $q$ and $c$, consider the request shape $\forall
  p . {\geqn 1q {\hasvalue(c)}}$.  The
  corresponding shape fragment is retrieved by the following
  SPARQL query:
{\sf \begin{tabbing}
SELECT ?s ?p ?o WHERE \{ \\
\{ \=SELECT ?v WHERE \\
   \>\{ \=?v $p$ ?x MINUS \{ ?v $p$ ?y \=OPTIONAL \{ ?y $q$ $c$ . ?v $p$ ?z \} \\
		                   \>\>\>FILTER (!bound(?z)) \} \} \} . \\
\{ \=\{ \=SELECT (?v AS ?s) ($p$ AS ?p) (?x as ?o) \\
      \>\>WHERE \{ ?v $p$ ?x . ?x $q$ $c$ \} \} \\
   \>UNION \\
   \>\{ \=SELECT (?x AS ?s) ($q$ AS ?p) ($c$ as ?o) \\
      \>\>WHERE \{ ?v $p$ ?x . ?x $q$ $c$ \} \} \} \}
\end{tabbing}}
The first subselect retrieves nodes $?v$ conforming to the shape;
the UNION of the next two subselects then retrieves the neighborhoods.
\qed
\end{example}

One may wonder about the converse to
Proposition~\ref{propshaclsparql}: is every SPARQL select-query
expressible as a shape fragment?  This does not hold, if only
because shape fragments always consist of triples from the input
graph, while select-queries can return arbitrary variable
bindings.  However, also more fundamentally, SHACL is strictly
weaker than SPARQL; we give two representative examples.

\begin{description}

\item[4-clique] Let $p \in I$.  There does not exist a shape
$\phi$ such that, on any RDF graph $G$, the nodes that conform to
$\phi$ are exactly the nodes belonging to a 4-clique of
$p$-triples in $G$.  We can show that if 4-clique would be
expressible by a shape, then the corresponding 4-clique query
about a binary relation $P$ would be expressible in 3-variable
counting infinitary logic $C^3_{\infty\omega}$.  The latter is
known not to be the case, however \cite{otto_bounded}.
(Infinitary logic is needed here to express path expressions, and
counting is needed for the $\mathord \geq_n$ quantifier, since we
have only 3 variables.)

\item[Majority] Let $p,q \in I$.  There does not exist a shape
$\phi$ such that, on any RDF graph $G$, the nodes that conform to
$\phi$ are exactly the nodes $v$ such that $\sharp \{x \mid
(v,p,x) \in G\} \geq \sharp \{x \mid (v,q,x) \in G\}$ (think of
departments with at least as many employees as projects).  We can
show that if Majority would be expressible by a shape, then the
classical Majority query about two unary relations $P$ and
$Q$ would be expressible in first-order logic.  Again, the latter
is not the case \cite{kolaitis_expressivepower}.
(Infinitary logic is not needed here, since for
this query, we can restrict to a class of structures where all
paths have length one.)
\end{description}

\subsection{Adapting a validation engine}

A shape fragments processor may also be obtained by adjusting a
SHACL validator to return the validated RDF terms and their
neighborhood, instead of a validation report.

A SHACL validation engine checks whether a given RDF graph
conforms to a given schema, and produces a validation report
detailing eventual violations.  A validation engine needs to
inspect the neighborhoods of nodes anyway.  Hence, it requires
only reasonably lightweight adaptations to produce, in addition
to the validation report, also the nodes and their neighborhoods
that validate the shapes graph, without introducing significant
overheads for tracing out and returning these neighborhoods, compared
to doing validation alone.  

To test this hypothesis, we extended the open-source,
free-license engine pySHACL \cite{pyshacl}.  This is a
main-memory engine and it achieves high coverage for the core
fragment of SHACL \cite{shacltestsuite}.  Written in Python, we
found it easy to make local changes to the code
\cite{paulson2007developers}; starting out with
4501 lines of code, 482 lines were changed, added or deleted.
     Our current implementation covers most of SHACL core with
     the exception of complex path expressions.  Our software,
     called \textbf{pySHACL-fragments}, is available open-source
     \cite{attachment}.

\subsection{Experiments}

We validated our approach by 
(i) assessing the correctness and practical
applicability of shape fragments; 
(ii) measuring the overhead of shape fragment extraction,
compared to mere validation, using our pySHACL-fragments
implementation; and (iii) testing the viability of computing shape
fragments by translation to SPARQL.
All RDF graphs, shapes and queries mentioned below,
along with scripts to download the datasets and run the
experiments, are publicly available \cite{attachment}.

\subsubsection{Applicability of shape fragments}

We \emph{simulated} a range of SPARQL queries by shape fragments.
Queries were taken from the SPARQL benchmarks BSBM
\cite{bizer2009berlin} and WatDiv \cite{alucc2014diversified}.
Unlike a shape fragment, a SPARQL select-query does not
return a subgraph but a set of variable bindings.
SPARQL construct-queries do return RDF graphs directly,
but not necessarily subgraphs.
Hence, we followed the methodology of modifying
SPARQL select-queries to construct-queries that
return all \emph{images} of the
pattern specified in the where-clause.

For tree-shaped basic
graph patterns, with given IRIs in the predicate position of
triple patterns, we can always simulate the
corresponding subgraph query by a shape fragment.  Indeed, a
typical query from the benchmarks retrieves nodes with some
specified properties, some properties of these properties, and so
on. For example, a slightly simplified WatDiv query, modified into a
subgraph query, would be the following. (To avoid clutter, we
forgo the rules of standard IRI syntax.)
{\sf
\begin{tabbing}
CONSTRUCT WHERE \{ \\
?v0 caption ?v1 . ?v0 hasReview ?v2 . ?v2 title ?v3 .\\
?v2 reviewer ?v6 . ?v7 actor ?v6 \}
\end{tabbing}}
(Here, \textsf{CONSTRUCT WHERE} is the
SPARQL notation for returning all images of a basic graph
pattern.)  We can express the above query as the fragment
for the following request shape:
\begin{multline*}
\sf
\geqn 1{caption}\top \wedge \geqn 1{hasReview}(\geqn 1{title}\top
\\
\sf {}
\land \geqn 1{reviewer} \geqn 1 {{actor}^-} \top)
\end{multline*}

Of course, patterns can involve various SPARQL operators, going
beyond basic graph patterns.  Filter conditions on property
values can be expressed as node tests in shapes;  optional
matching can be expressed using $\mathord \geq_0$ quantifiers.
For example, consider a simplified version of the pattern of a
typical BSBM query:
{\sf
\begin{tabbing}
?v text ?t . FILTER langMatches(lang(?t),``EN'') \\
OPTIONAL \{ ?v rating ?r \}
\end{tabbing}}
The images of this pattern can be retrieved using the shape
$$ \sf \geqn 1 {title} \test(lang=\textsf{``EN''}) \land \geqn 0 {rating}
\top. $$

Interestingly, the BSBM workload
includes a pattern involving a combination of optional matching and a negated
bound-condition to express absence of a certain property (a
well-known trick \cite{ag_expsparql,chile_sparql_pods}).
Simplified, this pattern looks as follows:
{\sf
\begin{tabbing}
?prod label ?lab . ?prod feature 870 \\
OPTIONAL \{ ?prod feature 59 . ?prod label ?var \} \\
FILTER (!bound(?var))
\end{tabbing}}
The images of this pattern can be retrieved using the shape
$$ \sf \geqn 1 {label} \top \land \geqn 1 {feature} \hasvalue(870)
\land \leqn 0 {feature} \hasvalue(59). $$

A total of 39 out of 46 benchmark queries, modified to return
subgraphs, could be simulated by shape fragments in this manner.
The remaining seven queries involved features not supported by
SHACL, notably, variables in the property position, or
arithmetic.

We have verified equality between the 39
SPARQL subgraph queries executed on the benchmark data, and the
corresponding shape fragments.  This experiment served as a 
correctness test of our system pySHACL-fragments.

\subsubsection{Extraction overhead} \label{subsuboverhead}

To measure the overhead of extracting shape fragments, compared
to doing validation alone, we compared execution times of
retrieving shape fragments using pySHACL-fragments, with
producing the corresponding validation report using pySHACL\@.
Here, we used the SHACL performance benchmark
\cite{schaffenrath2020benchmark} which consists of a 30-million
triple dataset known as the ``Tyrolean Knowledge Graph'',
accompanied by 58 shapes.  For this experiment we have only
worked with the five segments given by the first $N$ million
triples of the dataset, for $N=1,\dots,5$.

\begin{figure}
    \centering
    \includegraphics[width=\linewidth]{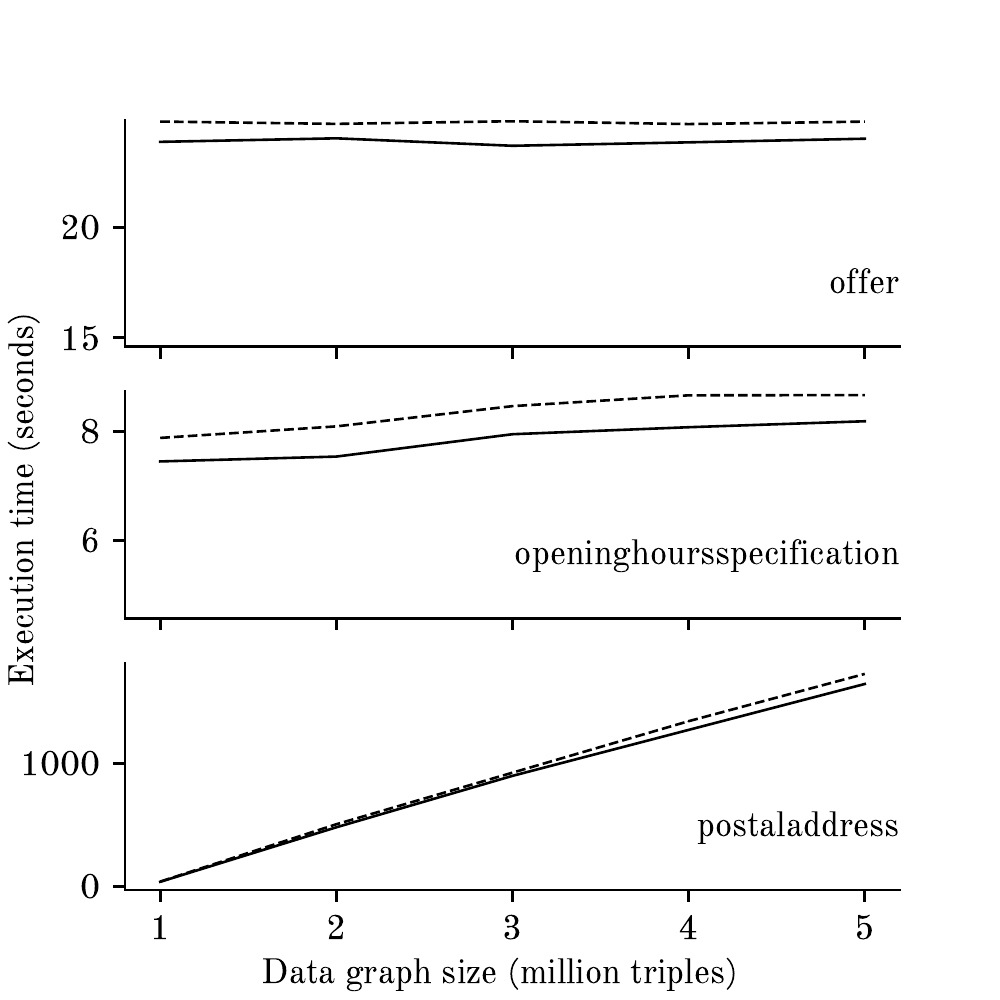}
    \caption{Adding shape extraction (dashed line) to pySHACL (full
    line) did not have large impact on the execution time, shown
    here for three representative
    shapes from the Tyrolean benchmark.}
    \label{fig:times}
\end{figure}

We executed each of the shapes five
times with each engine.  Timers were placed around the
\textsf{validator.run()} function, so \emph{data loading and shape
parsing time is not included.}  In this experiment, the average
overhead turns out to be below 10\%, as illustrated in
Figure~\ref{fig:times}.
We used a 2x 6core Intel Xeon E5-2620
v3s processor with 128GB DDR4 RAM and a 1TB hard disk.

\newcommand{\gelijk}[3]{\mathord=_{#1}{#2}.{#3}}
\newcommand{\gelet}[1]{\gelijk 1{#1}\top}

Going in more detail, we identified three different types of
behaviours.  Figure~\ref{fig:times} shows a representative plot
for each behaviour.  The first type of behavior shows a clear, linear,
increase in execution time for larger input sizes, going up to
thousands of seconds for size 5M.  This behavior occurs for three
benchmark shapes, among which the shape
\textsf{PostalAddressShape}.  We show this, and following
benchmark shapes, here in abridged form by
simplifying IRI notation and omitting
specific node tests.  Below we use
`$\tau$' to indicate the presence of a node test; we also
use $\gelijk nE\phi$ to abbreviate $\geqn nE\phi \land \leqn nE\phi$.

\begin{description}
  \item[PostalAddressShape:] \
\end{description}
\begin{multline*}
  \geqn 1{\mathsf{type}}{\hasvalue(\mathsf{PostalAddress})}
    \land \gelet {\mathsf{addressCountry}} \\
    {} \land \forall \mathsf{addressCountry}.\tau
    \land \gelet {\mathsf{addressLocality}}
    \land \forall \mathsf{addressCountry}.\tau \\
    {} \land \forall \mathsf{addressRegion}.\tau
    \land \geqn 1 {\mathsf{postalCode}} \top
    \land \forall \mathsf{postalCode}.\tau \\
    {} \land \gelet {\mathsf{streetAddress}}
    \land \forall \mathsf{streetAddress}.\tau
\end{multline*}

The second type of behavior shows only a modest increase in
execution time, increasing 10--20\% between sizes 1M and 5M. This
occurs for five benchmark shapes, among which the shape
\textsf{OpeningHourSpecificationShape}, shown below.  That the
slope of the linear increase is smaller here than in the previous
type can be explained by the distribution of nodes of type
Opening Hour Specification in the data segments, which occur less
densely than, e.g., Postal Addresses.

\begin{description}
  \item[OpeningHourSpecificationShape:] \
\end{description}
\begin{multline*}
  \geqn
  1{\mathsf{type}}{\hasvalue(\mathsf{OpeningHourSpecification})}
    \land \forall \mathsf{dayOfWeek}.\tau \\
    {} \land \forall \mathsf{closes}.\tau
    \land \forall \mathsf{opens}.\tau
    \land \forall \mathsf{validFrom}.\tau
    \land \forall \mathsf{validThrough}.\tau \\
    {} \land \leqn 1 {\mathsf{description}} \top
    \land \forall \mathsf{description}.\tau
\end{multline*}

The third and last type of behavior we observed shows execution
times that remain constant over the five data segments.  This
behavior actually occurs for the majority of the benchmark
shapes; we give \textsf{OfferShape} as an example.  The
explanation for this behavior is that all relevant
triples for these shapes already occur in the first segment of 1M
triples (recall that we do not measure data loading
time).  This first segment is indeed intended to be also used as
a self-contained benchmark dataset.

Independently of how execution times vary over the five data
segments, our measurements consistently report an average 10\%
overhead of extracting shape fragments.

\begin{description}
  \item[OfferShape:] \
\end{description}
\begin{multline*}
  \geqn 1{\mathsf{type}}{\hasvalue(\mathsf{Offer})}
    \land \gelet {\mathsf{name}}
    \land \forall \mathsf{name}.\tau
    \land \leqn 1{\mathsf{description}}\top \\
    {} \land \forall \mathsf{description}.\tau
    \land \gelet {\mathsf{availability}}
    \land \forall \mathsf{availability}.\tau \\
    {} \land \geqn 1{\mathsf{itemOffered}}\top
    \land \forall {\mathsf{itemOffered}}.(
  \geqn 1{\mathsf{type}}{\hasvalue(\mathsf{Service})} \lor {} \\
  \geqn 1{\mathsf{type}}{\hasvalue(\mathsf{Product})} \lor
  \geqn 1{\mathsf{type}}{\hasvalue(\mathsf{Apartment})}) \\
    {} \land \gelijk 1 {\mathsf{price}} \top
    \land \forall \mathsf{price}.\tau
    \land \gelijk 1 {\mathsf{priceCurrency}} \top
    \land \forall \mathsf{priceCurrency}.\tau \\
    {} \land \gelijk 1 {\mathsf{url}} \top
    \land \forall \mathsf{url}.\tau
    {} \land \forall \mathsf{validFrom}.\tau
    \land \forall \mathsf{validThrough}.\tau
\end{multline*}

\subsubsection{Computing shape fragments in SPARQL}

As already discussed in Section~\ref{secsparql}, shapes give
rise to complex SPARQL queries which pose quite a challenge to
SPARQL query processors.  It is outside the scope of the present
paper to do a performance study of SPARQL query processors; our
goal rather is to obtain an indication of the practical
feasibility of computing shape fragments in SPARQL\@.  Initial
work by Corman et al.\ has reported satisfying results on doing
\emph{validation} for nonrecursive schemas by a single, complex
SPARQL query \cite{shaclsparql}.  The question is whether we can
observe a similar situation
when computing shape fragments, where the
queries become even more complex.

We have obtained a mixed picture.  We used the main-memory SPARQL
engine Apache Jena ARQ\@.  Implementing the constructive proof of
Proposition~\ref{propshaclsparql}, we translated the shape
fragment queries for the benchmark shapes from the previous
Section~\ref{subsuboverhead} into large SPARQL queries.  The
generated expressions can be thousands of lines long, as our
translation procedure is not yet optimized to generate
``efficient'' SPARQL expressions.  However, we can reduce the
shapes by substituting $\top$ for node tests, and simplify the
resulting expressions.  For the three example shapes
\textsf{PostalAddressShape},
\textsf{OpeningHourSpecificationShape} and \textsf{OfferShape}
shown above, this amounts to substituting $\top$ for
$\tau$.  This reduction preserves the graph-navigational nature
of the queries.  Note also that, while a constraint like
$\forall p.\top$ is voidlessly true, it causes (as desired)
the inclusion of $p$-triples in the shape fragment.

\begin{figure}
    \centering
    \includegraphics[width=\linewidth]{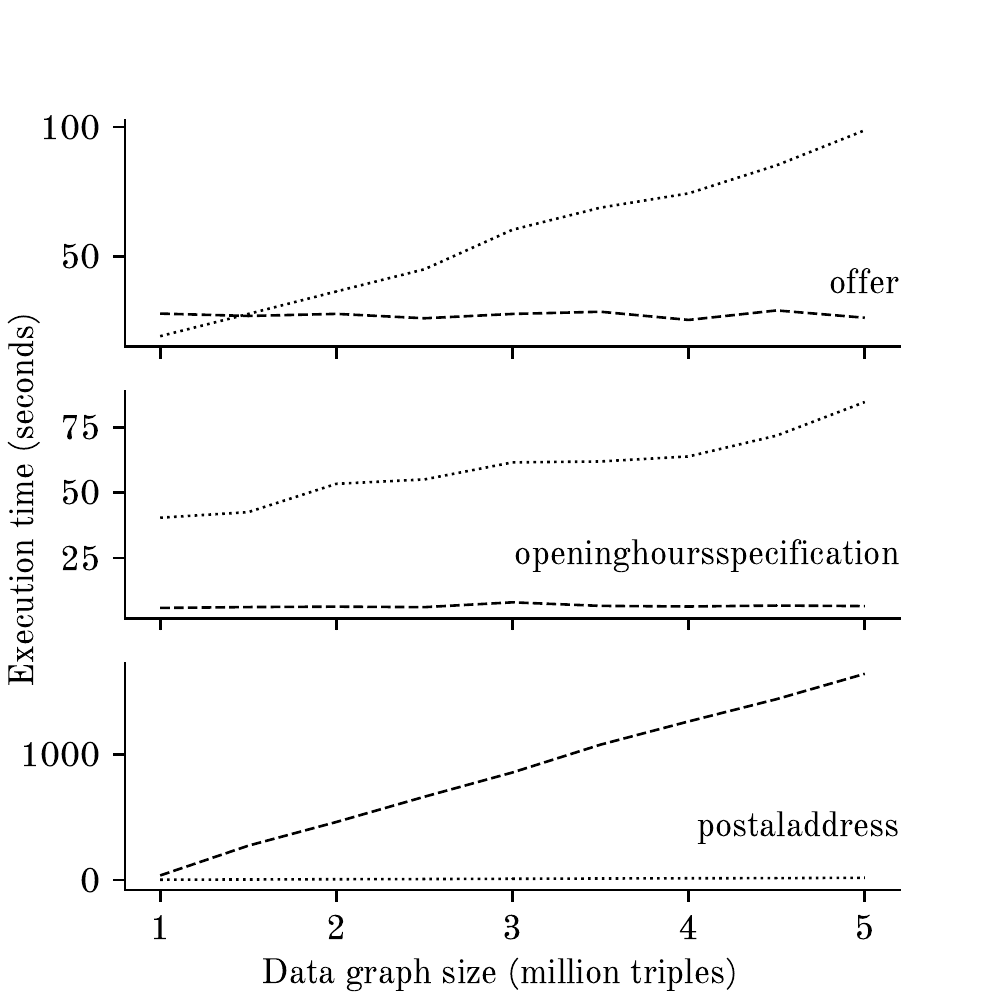}
    \caption{Jena ARQ in-memory SPARQL execution time (dotted) and pySHACL-fragments
    execution time (dashed).}
    \label{fig:sparqltimes}
\end{figure}

Execution times for the three SPARQL expressions, thus
simplified, are shown in Figure~\ref{fig:sparqltimes}, where they
are compared, on the same test data as before, with the
pySHACL-fragments implementation.  We realize this is an
apples-to-oranges comparison, but we can still draw some
tentatively positive conclusions.  Two SPARQL queries execute
slower than pySHACL-fragments, but not so much slower that a
log-scale $y$-axis would be needed to get them on the same
picture.  The SPARQL query for \textsf{PostalAddressShape} is
even much faster. This is explained by the absence of $\leq_n$
constraints, which have a complex neighborhood definition.  The
generated query has only joins and counts, but no negated
subqueries, which appears to run well on the ARQ processor.
Reported timings are averages over five runs.
We used a 2x 8core Intel Xeon E5-2650
v2 processor with 48GB DDR3 RAM and a 250GB hard disk.

Finally, to test the extraction of paths in SPARQL, we
used the DBLP database \cite{dblprdf}, and computed
the shape fragment for shape $\geqn
1{a^-/a/a^-/a/a^-/a}{\hasvalue(\mathsf{MYV})}$, where $a$ stands
for the property \textsf{dblp:authoredBy}, and \textsf{MYV}
stands for the DBLP IRI for Moshe Y.\ Vardi.  This fragment
extracts all authors at co-author distance three or less from
this famous computer scientist, plus all $a$-triples involved.
The generated SPARQL query is similar to the query from
Example~\ref{ex-qrpaths}.

\begin{figure}
    \centering
    \includegraphics[width=\linewidth]{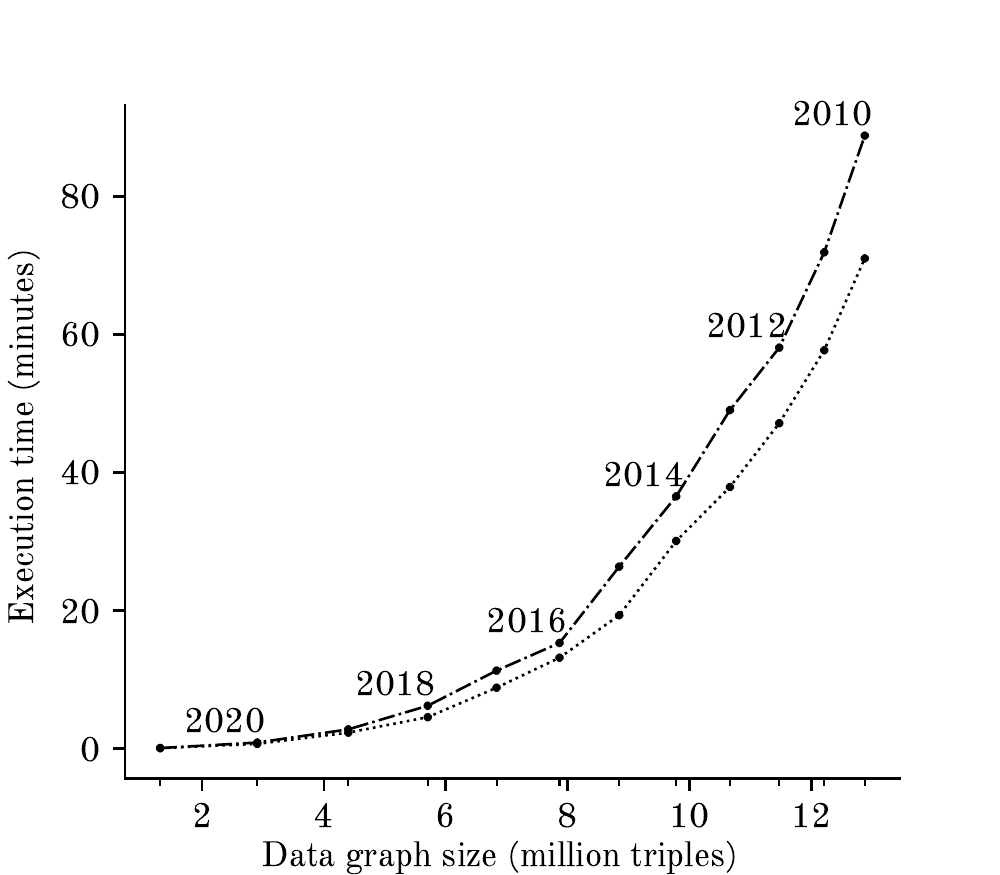}
    \caption{Jena ARQ store-based SPARQL execution time
    (dotted) and store-based GraphDB execution time (dashed-dotted)
    for the Vardi-distance-3 shape fragment.}
    \label{fig:varditimes}
\end{figure}

We ran this heavy analytical query on the two
secondary-memory engines Apache Jena ARQ on TDB2 store, and GraphDB\@.  The
execution times over increasing slices of DBLP, going backwards
in time from 2021 until 2010, are comparable between the two
engines (see Figure~\ref{fig:varditimes}).
Vardi is a prolific and central author and co-author; just from 2016
until 2021, almost 7\% of all DBLP authors are at distance
three or less, or almost $145\,943$ authors.
The resulting shape fragment contains almost 3\% of all
\textsf{dblpl:authoredBy} triples, or $219\,085$ unique triples.

\section{Related Work} \label{secrelated}

\subsection{Provenance}

Shapes may be viewed as queries on RDF graphs, returning the
nodes that conform to the shape.  A close affinity is then
apparent between neighborhoods as defined in this paper, and
notions of \emph{provenance} for database queries
\cite{provenance_whyhowwhere,glavic-data-provenance}.  

\newcommand{\pone}{\textsf{p1}}
\newcommand{\Anne}{\textsf{Anne}}
\newcommand{\Bob}{\textsf{Bob}}
\newcommand{\Paper}{\mathrm{Paper}}
\newcommand{\Author}{\mathrm{Author}}
\newcommand{\Student}{\mathrm{Student}}

A seminal work in the area of data provenance is that on
\emph{lineage} by Cui, Widom and Wiener \cite{lineage}.  Like
shape fragments, the lineage of a tuple returned by a query on a
database $D$ is a subdatabase of $D$.  Lineage was defined for
queries expressed in the relational algebra.
In principle, we can express shapes in relational
algebra.  So, instead of defining our own notion of neighborhood,
should we have simply used lineage instead?  The answer is no;
the following example shows that Theorem~\ref{theorem}
and Corollary~\ref{corol} would fail.

\begin{example} \label{exlineage}
Recalling Example~\ref{exleq},
consider a relational database schema with three relation schemes
$\Paper(P)$, $\Author(P,A)$, and $\Student(A)$,
and the integrity constraint that every paper should have at least one
author, but no paper should
have a non-student author.  Consider the database $D$ given by
$$ D(\Paper) = \{\pone\}; \
D(\Author) = \{(\pone,\Bob)\}; \
D(\Student) = \{\Bob\}. $$  Note that $D$
satisfies the integrity constraint.  The papers with at least one author
but without non-student authors are retrieved by the relational
algebra expression $Q = \Paper \Join (\pi_P(\Author) - V)$
  with $V = \pi_P(\Author - (\Author \Join \Student))$.
Since $V$ is empty on $D$,
the lineage of \pone\ for $Q$ in $D$ is the database $D'$ where
$$D'(\Paper) = \{\pone\}; \
D'(\Author) = \{(\pone,\Bob)\}; \
D'(\Student) = \emptyset. $$ This database
no longer satisfies the integrity constraint.
\qed
\end{example}

An alternative approach to lineage is \emph{why-provenance}
\cite{whywhere}.  This approach is non-deterministic in that it
reflects that there may be several ``explanations'' for why a
tuple is returned by a query (for example, queries involving
existential quantification).  Accordingly, why-provenance does
not yield a single neighborhood (called witness), but a set of
them.  While logical, this approach is at odds with our aim of
providing a \emph{deterministic} retrieval mechanism through
shapes.  Of course, one could take the union of all witnesses,
but this runs into similar problems as illustrated in the above
example.  Indeed, why-provenance was not developed for queries
involving negation or universal quantification.

A recent approach to provenance for negation is that by Gr\"adel
and Tannen \cite{gt_negation,tannen-siglog} based on the
successful framework of provenance semirings
\cite{provsemirings}.  There, provenance is produced in the form
of provenance polynomials which give a compact representation of
the several possible proof trees showing that the tuple satisfies
the query.  Thus, like why-provenance, this approach is
inherently non-deterministic.  Still, we were influenced by
Gr\"adel and Tannen's use of negation normal form, which we have
followed in this work.

\subsection{Triple pattern fragments}
Figure~\ref{fig:shapeFragments} is not meant to reflect
expressiveness comparisons.  That figure places shape fragments to the
right of triple pattern fragments (TPF \cite{tpf}).  However,
while shape fragments are in general obviously much more powerful than
TPFs, not all TPFs are actually expressible by shape fragments.

A TPF may be viewed as a query that, on an input graph $G$, returns
the subset of $G$ consisting of all images of some fixed triple
pattern in $G$.  For example, TPFs of the form $(?x,p,?y)$,
$(?x,p,c)$, $(c,p,?x)$, or $(c,p,d)$, for IRIs $p$, $c$, and $d$,
are easily expressed as shape fragments
using the request shapes $\geqn 1p\top$, $\geqn 1p\hasvalue(c)$,
$\geqn 1{p^-}\hasvalue(c)$, or
$\hasvalue(c) \land \geqn 1p{\hasvalue(d)}$,
respectively.

The TPF $(?x,p,?x)$, asking for all $p$-self-loops
in the graph, corresponds to the shape fragment for
$\neg \disj(\id,p)$.  A TPF of the form $(c,p,d)$, for IRI $d$,
asking for a single triple on condition that it is
present in the graph, corresponds to the shape fragment for

Furthermore,
the TPFs $(?x,?y,?z)$ (requesting a full download) and
$(c,?y,?z)$ are expressible using the request shapes
$\neg \closed(\emptyset)$ and
$\hasvalue(c) \land \neg \closed(\emptyset)$.  Here, the need to
use a ``trick'' via negation of closedness constraints exposes a
weakness of shapes: properties are not treated on equal footing
as subjects and objects.  Indeed, other TPFs involving variable
properties, such as $(?x,?y,c)$, $(?x,?y,?x)$, or $(c,?x,d)$, are
not expressible as shape fragments.

The above discussion can be summarized as follows.  The proof is
in \relegate.

\begin{proposition} \label{proptpfexp}
  The TPFs expressible as a shape fragment (uniformly over all
  input graphs) are precisely the TPFs of
  the following forms:
  \begin{enumerate}
    \item $(?x,p,?y)$;
\item $(?x,p,c)$;
    \item $(c,p,?x)$;
    \item $(c,p,d)$;
    \item $(?x,p,?x)$;
    \item $(?x,?y,?z)$;
    \item $(c,?y,?z)$.
\end{enumerate}
\end{proposition}

\begin{remark}
  SHACL does not allow \emph{negated properties} in path
expressions, while these are supported in SPARQL property paths.  
Extending SHACL with negated properties would readily allow the
  expression of \emph{all} TPFs as shape fragments.
  For example, the TPF $(?x,?y,c)$, for
  IRI $c$, would become expressible by requesting the shape
  $$ \geqn 1p{\hasvalue(c)} \lor \geqn
  1{!p}{\hasvalue(c)}, $$ with $p$ an arbitrary IRI.
  Here, the negated property $!p$ matches any property different from $p$.
\end{remark}

\subsection{Knowledge graph subsets}

Very recently, the idea of defining subgraphs (or fragments as we
call them) using shapes was independently proposed by Labra Gayo
\cite{labra-subsets}.  An important difference with our
SHACL-based approach is that his approach is based on ShEx, the
other shape language besides SHACL that is popular in practice
\cite{shex,validatingrdf}.  Shapes in ShEx are quite different
from those in SHACL, being based on bag-regular expressions over
the bag of properties of the focus node.  As a result, the
technical developments of our work and Labra Gayo's are quite
different.  Still, the intuitive and natural idea of forming a
subgraph by collecting all triples encountered during conformance
checking, is clearly the same in both approaches.  This idea,
which Labra Gayo calls ``slurping'', is implemented in our
pyshacl-fragments implementation, as well as a ``slurp'' option
in the shex.js implementation of ShEx \cite{shexjs}.  Labra Gayo
also gives a formal definition of ShEx + slurp, extending the
formal definition of ShEx \cite{shex}.

In our work we make several additional contributions compared to
the development by Labra Gayo:
\begin{itemize}
  \item
    We consider the important special case of
    shape fragments based on schemas with targets.
  \item
    We support path expressions directly, which in ShEx need to
    be expressed through recursion.
  \item
    We support negation, universal
    quantification, and other non-monotone quantifiers and
    shapes, such as $\leq_n$, equality, disjointness, lessThan.
  \item
    We make the connection to database provenance and to Linked
    Data Fragments.
  \item
    We establish formal correctness properties (the Conformance
    Theorem and the Sufficiency Lemma).
  \item
    We investigate the translation of shape fragments into
    SPARQL\@.  On the other hand, Labra Gayo discusses
    Pregel-based implementations of his query mechanism.
\end{itemize}

\subsection{Path-returning queries on graph databases}

Our definition of neighborhood of a node $v$ for a shape
involving a path expression $E$
returns $E$-paths from $v$ to relevant nodes $x$
(see Table~\ref{tabdefrag}).  Notably, these paths are returned
as a subgraph, using the $\graph$ constructor applied to a set of
paths.  Thus, shape fragments are loosely related to path-returning
queries on graph databases, introduced as a theoretical concept
by Barcel\'o et al.\ \cite{ecrpq-tods} and found in the languages Cypher
\cite{cypher} and G-CORE \cite{g-core}.

However, to our knowledge, a mechanism to return a set of paths
in the form of a subgraph is not yet implemented by these
languages.  We have showed in Section~\ref{secsparql} that, at
least in principle, this is actually possible in any standard
query language supporting path expressions, such as SPARQL\@.
Barcel\'o et al.\ consider a richer output structure whereby an
infinite set of paths (or even set of tuples of paths) resulting
from an extended regular path query can be finitely and
losslessly represented.  In contrast, our $\graph$ constructor is
lossy in that two different sets $S_1$ and $S_2$ of paths may
have $\graph(S_1)=\graph(S_2)$.  However, our Sufficiency Lemma
shows that our representation is sufficient for the purpose of
validating shapes.

\section{Conclusion} \label{seconcl}

The idea that shapes can serve not only for data validation, but
also for data access, has been floating largely informally within
the community (e.g., \cite{footprints,rubenshapeblog}).  Our work
is a step towards putting this idea on a solid formal footing.
Many questions open up for further research; we mention a few
very briefly.

SHACL is a quite powerful language, so an obvious direction is to
investigate efficient processing and optimization strategies for
SHACL, both just for validation, and for computing shape
fragments.  Recent work on validation optimization was done by
Figuera et al.\ \cite{travshacl}. Yet we believe many more
insights from database query optimization can be beneficial and
specialized to shape processing.  (A related direction is to use
shapes to inform SPARQL query optimization
\cite{SelectivityEstimation,ShapeStats}.)

We have seen that shape fragments
are strictly less expressive than SPARQL subgraph queries.  Is
the complexity of evaluation lower?  For those SPARQL subgraph
queries that \emph{are} expressible as shape fragments, are
queries in practice often easier to write in SHACL\@?
Can we precisely characterise the expressive power of SHACL?

Our approach to defining neighborhoods has been somehow
\emph{minimal} and \emph{deterministic}.  However, we miss
postulates stating in what precise sense our definitions (or
improved ones) are really minimal.

The SHACL recommendation only defines the semantics for
nonrecursive shape schemas, and we have seen in this paper that
shape fragments are already nontrivial for this case.
Nevertheless, there is current interest in \emph{recursive} shape schemas
\cite{shex,validatingrdf,shaclsparql,andresel,bj-recshacl}.  Extending
shape fragments with recursion is indeed another interesting
direction for further research.

Finally, it would be
desirable to extend shapes so that properties are treated
on equal footing as subjects and objects, as is indeed the spirit
of RDF \cite{nSPARQL,domagoj_trial_tods}.

\bibliographystyle{ACM-Reference-Format}
\bibliography{extra,database}

\onecolumn

\newgeometry{textwidth=40em}

\appendix

\section*{Appendix}

\newcommand{\shacl}{\textsc{SHACL}\xspace}
\newcommand{\schema}{\mathit{H}} 
\newcommand{\sgraph}{\mathcal{S}} 

\section{Translating real SHACL to formal SHACL}
\label{appreal2formal}

In this section we define the function $t$ which maps a SHACL shapes
graph $\sgraph$ to a schema $\schema$.

Assumptions about the shapes graph:
\begin{itemize}
\item All shapes of interest must be explicitly declared to be a
\texttt{sh:NodeShape} or \texttt{sh:PropertyShape}
\item The shapes graph is \href{https://www.w3.org/TR/shacl/\#dfn-well-formed}{well-formed}
\end{itemize}

Let the sets \(\sgraph_n\) and \(\sgraph_p\) be the sets of all node
shape shape names, respectively property shape shape names defined in
the shapes graph \(\sgraph\). Let \(d_x\) denote the set of RDF
triples in $\sgraph$ with \(x\) as the subject. We define $t(\sgraph)$
as follows:
$$
  t(\sgraph)=\{(x,t_{\mathit{nodeshape}}(d_x),
               t_{\mathit{target}}(d_x)) \mid x \in \sgraph_n\} \cup
\{(x,t_{\mathit{propertyshape}}(d_x), t_{\mathit{target}}(d_x)) \mid
               x \in \sgraph_p\}
$$
where we define $t_{\mathit{nodeshape}}(d_x)$ in Section \ref{sec:defin-t_nodeshape}, $t_{\mathit{propertyshape}}(d_x)$ in Section \ref{sec:defin-t_propertyshape} and $t_{\mathit{target}}(d_x)$ in Section \ref{sec:defin-t_target}.

\begin{remark}
We treat node shapes and property shapes separately. In particular, minCount, maxCount, qualified minCount, qualified maxCount, and uniqueLang constraints are only treated below under property shapes. Strictly speaking, however, these constraints may also be used in node shapes, where they are redundant, as the count equals one in this case.  For simplicity, we assume the shapes graph does not contain such redundancies. 
\end{remark}

\subsection{Defining $t_{\mathit{nodeshape}}(d_x)$}
\label{sec:defin-t_nodeshape}
This function translates SHACL node shapes to shapes in the
formalization. We define $t_{\mathit{nodeshape}}(d_x)$ to be the
following conjunction:
\begin{equation*}
   t_{\mathit{shape}}(d_x) \land t_{\mathit{logic}}(d_x) \land t_{\mathit{tests}}(d_x) \land t_{\mathit{value}}(d_x) \land t_{\mathit{in}}(d_x) \land t_{\mathit{closed}}(d_x) \land t_{\mathit{pair}}(\id, d_x) \land t_{\mathit{languagein}}(d_x)
\end{equation*}
where we define $t_{\mathit{shape}}(d_x)$, $t_{\mathit{logic}}(d_x)$,
$t_{\mathit{tests}}(d_x)$, $t_{\mathit{value}}(d_x)$,
$t_{\mathit{in}}(d_x)$, $t_{\mathit{closed}}(d_x)$,
$t_{\mathit{langin}}(d_x)$ and $t_{\mathit{pair}}(\id, d_x)$ in the
following subsections.

\subsubsection{Defining $t_{\mathit{shape}}(d_x)$}
\label{sec:defin-t_shape}
This function translates the
\href{https://www.w3.org/TR/shacl/\#core-components-shape}{Shape-based
  Constraint Components} from $d_x$ to shapes from the
formalization. This function covers the SHACL keywords:
\texttt{sh:node} and \texttt{sh:property}.

We define $t_{\mathit{shape}}(d_x)$ to be the conjunction:

\begin{equation*}
  \bigwedge_{(x, \mathtt{sh{:}node}, y)\in d_x} \hasshape(y) \land
  \bigwedge_{(x, \mathtt{sh{:}property}, y)\in d_x} \hasshape(y)
\end{equation*}

\subsubsection{Defining $t_{\mathit{logic}}(d_x)$}
\label{sec:defin-t_logic}
This function translates the
\href{https://www.w3.org/TR/shacl/\#core-components-logical}{Logical
  Constraint Components} from $d_x$ to shapes from the
formalization. This function covers the SHACL keywords:
\texttt{sh:and}, \texttt{sh:or}, \texttt{sh:not}, \texttt{sh:xone}.

We define $t_{\mathit{logic}}(d_x)$ as follows:
\begin{align*}
  & \bigwedge_{(x, \mathtt{sh{:}not}, y)\in d_x} (\neg \hasshape(y)) \land {} \\ 
  & \bigwedge_{(x, \mathtt{sh{:}and}, y)\in d_x} (\bigwedge_{z\in y} \hasshape(z)) \land {} \\
  & \bigwedge_{(x, \mathtt{sh{:}or}, y)\in d_x} (\bigvee_{z\in y} \hasshape(z)) \land {} \\
  & \bigwedge_{(x, \mathtt{sh{:}xone}, y)\in d_x}
    (\bigvee_{a \in y} (a \land \bigwedge_{b\in y - \{a\}} \neg \hasshape(b))) 
\end{align*}
where we note that the object $y$ of the triples with the predicate
$\mathtt{sh{:}and}$, $\mathtt{sh{:}or}$, or $\mathtt{sh{:}xone}$ is a
SHACL list.

\subsubsection{Defining $t_{\mathit{tests}}(d_x)$}
\label{sec:defin-t_tests}

This function translates the
\href{https://www.w3.org/TR/shacl/\#core-components-value-type}{Value
  Type Constraint Components},
\href{https://www.w3.org/TR/shacl/\#core-components-range}{Value Range
  Constraint Components}, and
\href{https://www.w3.org/TR/shacl/\#core-components-string}{String-based
  Constraint Components}, with exception to the \texttt{sh:languageIn}
keyword which is handled in Section \ref{sec:defin-t_langin}, from $d_x$
to shapes from the formalization. This function covers the SHACL
keywords:
\begin{quote}
\begin{raggedright}
\texttt{sh:class}, \texttt{sh:datatype}, \texttt{sh:nodeKind},
\texttt{sh:minExclusive}, \texttt{sh:maxExclusive},\\
\texttt{sh:minLength}, \texttt{sh:maxLength},
\texttt{sh:pattern}.
\end{raggedright}
\end{quote}

We define $t_{\mathit{tests}}(d_x)$ as follows:
\begin{align*}
  t_{\mathit{tests'}}(d_x) \land \bigwedge_{(x,
  \mathtt{sh{:}class}, y) \in d_x} \geq_1
  \mathtt{rdf{:}type}/\mathtt{rdf{:}subclassOf}^*. \hasvalue(y) 
\end{align*}
where $t_{\mathit{tests'}}(d_x)$ is defined next.  Let $\Gamma$
denote the set of keywords just mentioned above, except for
\texttt{sh:class}.
\begin{align*}
t_{\mathit{tests'}}(d_x) =
  \bigwedge_{c\in\Gamma} \bigwedge_{(x,c,y)\in d_x} \test(\omega_{c,y})
\end{align*}
where $\omega_{c,y}$
is the node test in $\Omega$ corresponding to the SHACL
constraint component corresponding to $c$ with parameter $y$. For
simplicity, we omit the $\mathtt{sh{:}flags}$ for
$\mathtt{sh{:}pattern}$.

\subsubsection{Defining $t_{\mathit{pair}}(\id, d_x)$}
\label{sec:defin-t_pairid}
This function translates the
\href{https://www.w3.org/TR/shacl/\#core-components-property-pairs}{Property
  Pair Constraint Components} when applied to a node shape from $d_x$
to shapes from the formalization. This function covers the SHACL
keywords: \texttt{sh:equals}, \texttt{sh:disjoint},
\texttt{sh:lessThan}. \texttt{sh:lessThanOrEquals}.

We define the function $t_{pair}(\id, d_x)$ as follows:
\begin{itemize}
\item If $\exists p: (x, \mathtt{sh{:}lessThan}, p)\in d_x$ or
  $(x, \mathtt{sh{:}lessThanEq}, p)\in d_x$, then we define
  $t_{pair}(\id, d_x)$ as $\bot$.
\item Otherwise, we define $t_{pair}(\id, d_x)$ as
  \[\bigwedge_{(x, \mathtt{sh{:}equals}, p) \in d_x} \eq(\id, p) \land
  \bigwedge_{(x, \mathtt{sh{:}disjoint}, p) \in d_x} \disj(\id, p)\]
\end{itemize}

\subsubsection{Defining $t_{\mathit{languagein}}(d_x)$}
\label{sec:defin-t_langin}
This function translates the constraint component
\href{https://www.w3.org/TR/shacl/#LanguageInConstraintComponent}{Language
  In Constraint Component} from $d_x$ to shapes from the
formalization. This function covers the SHACL keyword:
\texttt{sh:languageIn}.

The function $t_{\mathit{languagein}}(E,d_x)$ is defined as follows:
\begin{align*}
  t_{\mathit{languagein}}(E, d_x) = \bigwedge_{(x, \mathtt{sh{:}languageIn}, y) \in d_x} \forall E . \bigvee_{\mathit{lang} \in y} test(\omega_{\mathit{lang}})
\end{align*}
where $y$ is a SHACL list and $\omega_{\mathit{lang}}$ is the element
from $\Omega$ that corresponds to the test that checks if the node is
annotated with the language tag $\mathit{lang}$.

\subsubsection{Defining other constraint components}
\label{sec:defin-t_other}

These functions translate the
\href{https://www.w3.org/TR/shacl/\#core-components-others}{Other
  Constraint Components} from $d_x$ to shapes from the
formalization. This function covers the SHACL keywords:
\texttt{sh:closed}, \texttt{sh:ignoredProperties},
\texttt{sh:hasValue}, \texttt{sh:in}.

We define the following functions:
\begin{equation*}
  t_{\mathit{value}}(d_x) = \bigwedge_{(x, \mathtt{sh{:}hasValue}, y)\in d_x} \hasvalue(y)
\end{equation*}

\begin{equation*}
  t_{\mathit{in}}(d_x) = \bigwedge_{(x, \mathtt{sh{:}in}, y)\in d_x} (\bigvee_{a\in y} \hasvalue(a))
\end{equation*}

Let $P$ be the set of all properties $p\in I$ such that
$(y, \mathtt{sh{:}path}, p)\in \sgraph$ where $y$ is a property shape
such that $(x, \mathtt{sh{:}property}, y) \in d_x$ union the set given
by the SHACL list specified by the $\mathtt{sh{:}ignoredProperties}$
parameter. Then, we define the function $t_{\mathit{closed}}(d_x)$ as
follows:
\begin{equation*}
t_{\mathit{closed}}(d_x) =
\begin{cases}
  \top & \text{if $(x, \mathtt{sh{:}closed}, \mathit{true}) \not\in d_x$} \\
  \closed(P) & \text{otherwise}
\end{cases}
\end{equation*}

\newcommand{\pepe}{\mathit{pp}}
\subsection{Defining $t_{\mathit{path}}(\pepe)$}
\label{sec:defin-t_path}

In preparation of the next Subsection, 
this function translates the
\href{https://www.w3.org/TR/shacl/\#property-paths}{Property Paths} to
path expressions. This part of the translation deals with the SHACL
keywords:
\begin{quote}
\begin{raggedright}
\texttt{sh:inversePath},
\texttt{sh:alternativePath}, \texttt{sh:zeroOrMorePath},\\
\texttt{sh:oneOrMorePath},
\texttt{sh:zeroOrOnePath}, \texttt{sh:alternativePath}.
\end{raggedright}
\end{quote}

For an IRI or blank node $\pepe$ representing a property path, we define $t_{\mathit{path}}(\pepe)$ as follows:
\newcommand{\tpath}{t_{\mathit{path}}}
\begin{equation*}
  t_{\mathit{path}}(\pepe) =
  \begin{cases}
    \pepe & \text{if $\pepe$ is an IRI} \\
    t_{\mathit{path}}(y)^- & \text{if}\, \exists y: (\pepe, \mathtt{sh{:}inversePath}, y) \in \sgraph \\
    t_{\mathit{path}}(y)^* & \text{if}\, \exists y: (\pepe, \mathtt{sh{:}zeroOrMorePath}, y) \in \sgraph \\
    t_{\mathit{path}}(y) \comp t_{\it path}(y)^* & \text{if}\, \exists y: (\pepe, \mathtt{sh{:}oneOrMorePath}, y) \in \sgraph \\ 
    t_{\mathit{path}}(y)? & \text{if}\, \exists y: (\pepe, \mathtt{sh{:}zeroOrOnePath}, y) \in \sgraph \\ 
    \bigcup_{a\in y} t_{\it path}(a) &
\text{if $\exists y: (\pepe, \mathtt{sh{:}alternativePath}, y) \in
\sgraph$ and $y$ is a \shacl list} \\ 
    t_{\mathit{path}}(a_1) \comp \dots \comp t_{\mathit{path}}(a_n) &
    \text{if $\pepe$ represents the \shacl list $[a_1, \dots, a_n]$} 
  \end{cases}
\end{equation*}

\subsection{Defining $t_{\mathit{propertyshape}}(d_x)$}
\label{sec:defin-t_propertyshape}
This function translates SHACL property shapes to shapes in the
formalization. Let $\pepe$ be the property path associated with $d_x$. Let
$E$ be $t_{\mathit{path}}(\pepe)$. We define
$t_{\mathit{propertyshape}}(d_x)$ as the following conjunction:
\begin{equation*}
  t_{\mathit{card}}(E, d_x) \land
  t_{\mathit{pair}}(E, d_x) \land t_{\mathit{qual}}(E, d_x) \land
  t_{\mathit{all}}(E, d_x) \land t_{\mathit{uniquelang}}(E, d_x)
\end{equation*}
where we define $t_{\mathit{card}}$, 
$t_{\mathit{pair}}$,
$t_{\mathit{qual}}$,
$t_{\mathit{all}}$, and
$t_{\mathit{uniquelang}}$ in the following subsections.

\subsubsection{Defining $t_{\mathit{card}}(E, d_x)$}
\label{sec:defin-t_card}
This function translates the
\href{https://www.w3.org/TR/shacl/\#core-components-count}{Cardinality
  Constraint Components}. from $d_x$ to shapes from the
formalization. This function covers the SHACL keywords:
\texttt{sh:minCount}, \texttt{sh:maxCount}.

We define the function $t_{card}(E, d_x)$ as follows:

\begin{equation*}
  \bigwedge_{(x,\mathtt{sh{:}minCount}, n) \in d_x} \geq_n E.\top \land \bigwedge_{(x,\mathtt{sh{:}maxCount}, n) \in d_x} \leq_n E.\top 
\end{equation*}

\subsubsection{Defining $t_{\mathit{pair}}(E, d_x)$}
\label{sec:defin-t_pair}
This function translates the
\href{https://www.w3.org/TR/shacl/\#core-components-property-pairs}{Property
  Pair Constraint Components} when applied to a property shape from
$d_x$ to shapes from the formalization. This function covers the SHACL
keywords: \texttt{sh:equals}, \texttt{sh:disjoint},
\texttt{sh:lessThan}. \texttt{sh:lessThanOrEquals}.

We define the function $t_{pair}(E, d_x)$ as follows:
\begin{align*}
  & \bigwedge_{(x, \mathtt{sh{:}equals}, p) \in d_x} \eq(E, p) \land {}\\
  & \bigwedge_{(x, \mathtt{sh{:}disjoint}, p) \in d_x} \disj(E, p) \land {}\\
  & \bigwedge_{(x, \mathtt{sh{:}lessThan}, p) \in d_x} \lessthan(E, p) \land {}\\
  & \bigwedge_{(x, \mathtt{sh{:}lessThanOrEquals}, p) \in d_x} \lessthaneq(E, p) 
\end{align*}

\subsubsection{Defining $t_{\mathit{qual}}(E, d_x)$}
\label{sec:defin-t_qual}
This function translates the
\href{https://www.w3.org/TR/shacl/\#QualifiedValueShapeConstraintComponent}{(Qualified)
  Shape-based Constraint Components} from $d_x$ to shapes from the
formalization. This function covers the SHACL keywords:
\begin{quote}
\begin{raggedright}
\texttt{sh:qualifiedValueShape}, \texttt{sh:qualifiedMinCount},
\texttt{sh:qualifiedMaxCount},\\ \texttt{sh:qualifiedValueShapesDisjoint}.
\end{raggedright}
\end{quote}

We distinguish between the case where the parameter
$\mathtt{sh{:}qualifiedValueShapesDisjoint}$ is set to \textit{true},
and the case where it is not:
\begin{equation*}
t_{\it qual}(E, d_x) =
\begin{cases}
t_{\it sibl}(E, d_x) & \text{if $(x, \mathtt{sh{:}qualifiedValueShapesDisjoint}, \mathit{true}) \in d_x$} \\
t_{\it nosibl}(E, d_x) & \text{otherwise}
\end{cases}
\end{equation*}
where we define $t_{\it sibl}(E, d_x)$ and $t_{\it nosibl}(E, d_x)$ next.  Let
$\mathit{ps} = \{v \mid (v, \mathtt{sh{:}property}, x) \in
\sgraph\}$. We define the set of
\href{https://www.w3.org/TR/shacl/#dfn-sibling-shapes}{sibling shapes}
$$\mathit{sibl} = \{w \mid \exists v\in \mathit{ps}\, \exists y (v,
\mathtt{sh{:}property}, y):
(y, \mathtt{sh{:}qualifiedValueShape}, w) \in \sgraph\}.$$
We also define:
\newcommand{\kuu}{\mathcal Q}
\newcommand{\minkuu}{\kuu\mathit{min}}
\newcommand{\maxkuu}{\kuu\mathit{max}}
\begin{align*}
\kuu &= \{y \mid (x, \mathtt{sh{:}qualifiedValueShape}, y) \in d_x\} \\
\minkuu &= \{z \mid (x, \mathtt{sh{:}qualifiedMinCount}, z) \in
d_x\} \\
\maxkuu &= \{z \mid (x, \mathtt{sh{:}qualifiedMaxCount}, z) \in
d_x\}
\end{align*}
We now define
\begin{multline*}
t_{\mathit{sibl}}(E, d_x) =
\bigwedge_{y \in \kuu} \bigwedge_{z \in \minkuu}
  \geq_z E.(\hasshape(y) \land
  \bigwedge_{s \in \mathit{sibl}} \neg \hasshape(s))
  \\
{} \land
\bigwedge_{y \in \kuu} \bigwedge_{z \in \maxkuu}
  \leq_z E.(\hasshape(y) \land
  \bigwedge_{s \in \mathit{sibl}} \neg \hasshape(s)) 
\end{multline*}
and $$ t_{\mathit{nosibl}}(E, d_x) =
\bigwedge_{y \in \kuu} \bigwedge_{z \in \minkuu}
  \geq_z E.\hasshape(y)
\land
\bigwedge_{y \in \kuu} \bigwedge_{z \in \maxkuu}
  \leq_z E.\hasshape(y). $$

\subsubsection{Defining $t_{\mathit{all}}(E, d_x)$}
\label{sec:defin-t_all}
This function translates the constraint components that are not
specific to property shapes, but which are applied on property shapes.

We define the function $t_{\mathit{all}}(E, d_x)$ to be:
\begin{align*}
  \forall E.( t_{\mathit{shape}}(d_x) \land t_{\mathit{logic}}(d_x) \land t_{\mathit{tests}}(d_x) \land t_{\mathit{in}}(d_x) \land t_{\mathit{closed}}(d_x) \land t_{\mathit{languagein}}(d_x)) \land t_{\mathit{allvalue}}(E, d_x)
\end{align*}
where
\begin{equation*}
  t_{\mathit{allvalue}}(E,d_x) = 
  \begin{cases}
    \top & \text{if $\not\exists v: (x, \mathtt{sh{:}hasValue}, v)\in d_x$}\\
    \geqn{1}{E}{t_{\mathit{value}}(d_x)} & \text{otherwise}
  \end{cases}
\end{equation*}
and 
$t_{\mathit{shape}}$,
$t_{\mathit{logic}}$,
$t_{\mathit{tests}}$,
$t_{\mathit{value}}$,
$t_{\mathit{languagein}}$,
and $t_{\mathit{closed}}$ are as defined earlier.
Note the treatment of the
$\mathtt{sh{:}hasValue}$ parameter when used in a property
shape. Unlike the other definitions, it is not universally quantified
over the value nodes given by $E$.

\subsubsection{Defining $t_{\mathit{uniquelang}}(E, d_x)$}
\label{sec:defin-t_uniquelang}
This function translates the constraint component
\href{https://www.w3.org/TR/shacl/#UniqueLangConstraintComponent}{Unique
  Lang Constraint Component} from $d_x$ to shapes from the
formalization. This function covers the SHACL keyword:
\texttt{sh:uniqueLang}.

The function $t_{\mathit{uniquelang}}(E,d_x)$ is defined as follows:
\begin{equation*}
  t_{\mathit{uniquelang}}(E, d_x) =
  \begin{cases}
    \uniquelang(E) & \text{if $(x,\mathtt{sh{:}uniqueLang}, \mathit{true}) \in d_x$} \\
    \top & \text{otherwise}
  \end{cases}
\end{equation*}

\subsection{Defining $t_{\mathit{target}}(d_x)$}
\label{sec:defin-t_target}
This function translates the
\href{https://www.w3.org/TR/shacl/\#targets}{Target declarations} to
shapes from the formalization. This function covers the SHACL
keywords:
\begin{quote}
\texttt{sh:targetNode}, \texttt{sh:targetClass},
\texttt{sh:targetSubjectsOf}, \texttt{sh:targetObjectsOf}.
\end{quote}

We define the function as follows:
\begin{align*}
  t_{\mathit{target}}(d_x) = & \bigvee_{(x, \mathtt{sh{:}targetNode}, y) \in d_x} \hasvalue(y) \lor{} \\
  & \bigvee_{(x, \mathtt{sh{:}targetClass}, y)\in d_x} \geqn{1}{\mathtt{rdf{:}type}/\mathtt{rdf{:}subclassOf}^*}{\hasvalue(y)} \lor{}\\
  & \bigvee_{(x, \mathtt{sh{:}targetSubjectsOf}, y)\in d_x} \geqn{1}{y}{\top} \lor{}\\
  & \bigvee_{(x, \mathtt{sh{:}targetObjectsOf}, y)\in d_x} \geqn{1}{y^-}{\top}
\end{align*}
If none of these triples are in $d_x$ we define $t_{\mathit{target}}(d_x) = \bot$

\section{Proof of Conformance Theorem} \label{appthm}

Assuming the Sufficiency Lemma, the proof of the Conformance
Theorem straightforwardly goes as follows.  Let $F=\frag(G,H)$;
we must show that $F$ conforms to $H$.  Thereto, consider a shape
definition $(s,\phi,\tau) \in H$, and let $v$ be a node such that
$F,v \models \tau$.  Since $F \subseteq G$ and $\tau$ is
monotone, also $G,v \models \tau$, whence $G,v \models \phi$
since $G$ conforms to $H$.  Since by definition, $F$ contains
$B(v,G,\phi)$, the Sufficiency Lemma yields $F,v \models \phi$ as
desired.

Toward a proof of the Sufficiency Lemma, we first prove:

\begin{proof}[Proof of Proposition~\ref{prop:graphpaths}]
That $\iexpr EF \subseteq \iexpr EG$ is immediate from $F
\subseteq G$ and the monotonicity of path expressions.  For the
reverse inclusion, we proceed by induction on the structure of
$E$.  The base case, where $E$ is a property $p$, is immediate
from the definitions.  The inductive cases where $E$ is one of
$E_1 \cup E_2$, $E_1^-$, or $E_1/E_2$, are clear.  Two cases
remain:
  \begin{itemize}
  \item $E$ is $E_1?$.  If $a=b$, then $(a,b)\in \iexpr{E}{F}$ by
    definition. Otherwise, $(a,b)$ must be in
    $\iexpr{E_1}{G}$. Therefore, by induction,
    $(a,b) \in \iexpr{E_1}{F} \subseteq \iexpr{E}{F}$.
  \item $E$ is $E_1^*$. If $a=b$, then $(a,b)\in \iexpr{E}{F}$ by
    definition. Otherwise, there exist $i \geq 1$ nodes
    $c_0,\dots,c_i$ such that $a=c_0$ and $b=c_i$, and
    $(c_j,c_{j+1}) \in \iexpr{E_1}{G}$ for $0\leq j <
    i$. By induction, each
    $(c_j,c_{j+1}) \in \iexpr{E_1}{F}$, whence
    $(a,b)$ belongs to the transitive closure of $\iexpr{E_1}{F}$
    as desired.
  \end{itemize}
\end{proof}

We now give the

\begin{proof}[Proof of the Sufficiency Lemma]
  For any shape $\phi$, we consider its expansion with relation to the
  schema $H$, obtained by repeatedly replacing subshapes of the form
  $\hasshape(s)$ by $\sdef{s}{H}$, until we obtain an equivalent shape
  that no longer contains any subshapes of the form
  $\hasshape(s)$. The proof proceeds by induction on the height of the
  expansion of $\phi$ in negation normal form, where the height of
  negated atomic shapes is defined to be zero. When $\phi$ is $\top$,
  $\test(t)$, or $\hasvalue(c)$, and $v$ conforms to $\phi$ in $G$,
  then $v$ clearly also conforms to $\phi$ in $G'$, as the conformance
  of the node is independent of the graph. We consider the following
  inductive cases:
  \begin{itemize}
  \item $\phi$ is $\phi_1 \land \phi_2$. By induction, we know $v$
    conforms to $\phi_1$ in $G'$ and conforms to $\phi_2$ in
    $G'$. Therefore, $v$ conforms to $\phi_1 \land \phi_2$ in $G'$.
  \item $\phi$ is $\phi_1 \lor \phi_2$. We know $v$ conforms to at
    least one of $\phi_i$ for $i\in \{1,2\}$ in $G$. Assume
    w.l.o.g. that $v$ conforms to $\phi_1$ in $G$. Then, our claim
    follows by induction.
    
  \item $\phi$ is $\geqn{n}{E}\psi$. Here, and in the following cases,
    we denote $B(v,G,\phi)$ by $B$. As $G,v\models\phi$, we know there
    are at least $n$ nodes $x_1,\dots,x_n$ in $G$ such that
    $x_i \in \iexpr{E}{G}(v)$ and $G,x_i\models\psi$ for all
    $1 \leq i \leq n$. Let $F=\graph(\paths(E,G,v,x))$. By
    Proposition~\ref{prop:graphpaths}, $x_i\in\iexpr{E}{F}(v)$. By
    definition of $\phi$-neighborhood $F\subseteq B$, and we know
    $B\subseteq G'$. Therefore, because $E$ is monotone,
    $x_i\in\iexpr{E}{G'}(v)$. Furthermore, since
    $B(x_i,G,\psi) \subseteq B \subseteq G'$, by induction,
    $G',x_i\models \psi$ as desired.
    
  \item $\phi$ is $\leqn{n}{E}\psi$. First we show that every
    $x \in \iexpr{E}{G'}(v)$ that conforms to $\psi$ in $G'$, must also
    conform to $\psi$ in $G$.

    Proof by contradiction. Suppose there exists a node
    $x \in \iexpr{E}{G'}(v)$ that conforms to $\psi$ in $G'$, but
conforms to $\neg\psi$ in $G$. By definition of
    $\phi$-neighborhood, $B(x,G,\neg \psi)\subseteq B$, and we know
    $B \subseteq G'$. Therefore, by induction, $x$ conforms to
    $\neg \psi$ in $G'$, which is a contradiction.

    Because of the claim above, the number of nodes reachable from
    $v$ through $E$ that satisfy $\psi$ in $G'$ must be smaller or
    equal to the number of nodes reachable from $v$ through $E$ that
    satisfy $\psi$ in $G$. Because we know $G,v\models \phi$, the
    lemma follows. 

  \item $\phi$ is $\forall E.\psi$. For all nodes $x$ such that
    $x\in\iexpr{E}{G'}(v)$, as $E$ is monotone,
    $x\in\iexpr{E}{G}(v)$. As $G,v\models\phi$, $G,x\models\psi$. By
    definition of $\phi$-neighborhood, $B(x,G,\psi) \subseteq B$. We
    know $B \subseteq G'$. Thus, by induction, $G',x\models\psi$ as
    desired.
    
  \item $\phi$ is $\eq(E,p)$. We must show that
    $\iexpr{E}{G'}(v) = \iexpr{p}{G'}(v)$. For the containment from
    left to right, let $x\in\iexpr{E}{G'}(v)$. Since $E$ is monotone,
    $x\in\iexpr{E}{G}(v)$. Since $G,v\models\phi$,
    $x\in\iexpr{p}{G}(v)$. Let $F=\graph(\paths(p,G,v,x))$. By
    Proposition~\ref{prop:graphpaths}, $x\in\iexpr{p}{F}(v)$. By
    definition of $\phi$-neighborhood, $F\subseteq B$, and we know
    $B\subseteq G'$. Therefore, because path expressions are monotone,
    we also have $x\in\iexpr{p}{G'}(v)$ as desired. The containment
    from right to left is analogous.

  \item $\phi$ is $\eq(\id, p)$. We must show that
    $\iexpr{id}{G'}(v) = \iexpr{p}{G'}(v)$, or equivalently we must
    show that $\{v\} = \iexpr{p}{G'}(v)$. We know that
    $G,v\models \phi$, therefore $\iexpr{p}{G}(v) = \{v\}$. Now we
    only need to show that $(v,p,v)\in G'$ as $G'\subseteq G$ (and
    therefore $G'$ does not contain more $p$-edges than $G$). Because
    by definition of neighborhood $B = \{(v,p,v)\}$, and because
    $B\subseteq G'$, the claim follows.
    
  \item $\phi$ is $\disj(E,p)$. Let $x\in\iexpr{E}{G'}(v)$. Since $E$
    is monotone, $x\in\iexpr{E}{G}(v)$. Since $G,v\models \phi$,
    $x\not\in\iexpr{p}{G}(v)$. Therefore, as $p$ is monotone,
    $x\not\in\iexpr{p}{G'}(v)$. The case where $x\in\iexpr{p}{G'}(v)$ is
    analogous.

  \item $\phi$ is $\disj(\id,p)$. We must show that
    $(v,p,v)\not\in G'$. Because $G,v\models \phi$, we know that
    $(v,p,v)\not\in G$. As $G'\subseteq G$, the claim follows.
    
  \item $\phi$ is $\lessthan(E,p)$. Let $x \in \iexpr{E}{G'}(v)$. Let
    $(v,p,y) \in G'$. We must show that $x < y$. Since $E$ is
    monotone, $x\in\iexpr{E}{G}(v)$ and since $G'\subseteq G$,
    $(v,p,y) \in G$. As $G,v\models \phi$, we know that $x < y$ in $G$
    and thus also in $G'$.
    
  \item $\phi$ is $\lessthaneq(E,p)$. Analogous to the case where
    $\phi$ is $\lessthan(E,p)$.
    
  \item $\phi$ is $\uniquelang(E)$. Let $x\in\iexpr{E}{G'}(v)$. Let
    $y\in\iexpr{E}{G'}(v)$ such that $y\neq x$. As $E$ is monotone,
    $x\in\iexpr{E}{G}(v)$ and $y\in\iexpr{E}{G}(v)$. As
    $G,v\models\phi$, we know $y \nsim x$ in $G$ and thus also in
    $G'$.
    
  \item $\phi$ is $\closed(P)$. Let $(v,p,x) \in G'$. Then,
    $(v,p,x) \in G$. Therefore, as $G,v\models\phi$, $p\in P$ as
    desired.
    
  \item $\phi$ is $\neg \eq(E,p)$. Since $G,v\models\phi$, there are
    two cases. First, there exists a node $x\in\iexpr{E}{G}(v)$ such
    that $x\not\in\iexpr{p}{G}(v)$. Let
    $F=\graph(\paths(E,G,v,x))$. By Proposition~\ref{prop:graphpaths},
    $x\in\iexpr{E}{F}(v)$. By definition of $\phi$-neighborhood
    $F\subseteq B$, and we know $B\subseteq G'$. Therefore, since $E$
    is monotone, $x\in \iexpr{E}{G'}(v)$. Next, since
    $(v,p,x)\not\in G$, we know $(v,p,x)\not\in G'$. Thus,
    $\iexpr{E}{G}(v) \neq \iexpr{p}{G}(v)$ as desired. For the other
    case, there exists a node $x$ such that $(v,p,x)\in G$ and
    $x\not\in \iexpr{E}{G}(v)$. By definition of $\phi$-neighborhood,
    $(v,p,x)\in B\subseteq G'$. However, because $E$ is monotone
    $x\not\in \iexpr{E}{G'}(v)$. Therefore
    $\iexpr{p}{G}(v)\neq \iexpr{E}{G}(v)$ as desired.

  \item $\phi$ is $\neg \eq(\id, p)$. Since $G,v\models\phi$, there
    are two cases. First, $(v,p,v)\not\in G$. We know $G'\subseteq G$,
    therefore if $(v,p,v)\not\in G$, then $(v,p,v)\not\in G'$ as
    desired. Second, $(v,p,v)\in G$ and there exists a node $x$ such
    that $(v,p,x)\in G$ and $x\neq v$. From the definition of
    neighborhood, we know that this also holds for $B$ and therefore
    also in $G'$ as $B\subseteq G'$.
    
  \item $\phi$ is $\neg \disj(E,p)$. Since $G,c\models\phi$, we know
    that there exists a node $x\in\iexpr{E}{G}(v)$ such that
    $(v,p,x)\in G$. Let $F=\graph(\paths(E,G,v,x))$. By
    Proposition~\ref{prop:graphpaths}, $x\in\iexpr{E}{F}(v)$. By
    definition of $\phi$-neighborhood $F\subseteq B$, and we know
    $B\subseteq G'$. Then, since $E$ is monotone,
    $x\in \iexpr{E}{G'}(v)$. Next, by definition of
    $\phi$-neighborhood, also $(v,p,x)\in B\subseteq G'$. Thus,
    $x\in \iexpr{E}{G'}(v) \cap \iexpr{p}{G'}(v)$ as desired.

  \item $\phi$ is $\neg \disj(\id,p)$. We need to show that
    $(v,p,v)\in G'$. By definition of neighborhood, $(v,p,v)\in B$. As
    $B\subseteq G'$, $(v,p,v)\in G'$ as desired.
    
  \item $\phi$ is $\neg \lessthan(E,p)$. Since $G,v\models\phi$, there
    exists a node $x\in\iexpr{E}{G}(v)$ and a node
    $y\in\iexpr{p}{G}(v)$ with $x \nless y$. If we can show that
    $x\in\iexpr{E}{G'}(v)$ and $x\in\iexpr{p}{G'}(v)$, it will follow
    that $G',v\models\phi$ as desired.  Let
    $F=\graph(\paths(E,G,v,x))$. By Proposition~\ref{prop:graphpaths},
    $x\in\iexpr{E}{F}(v)$. By definition of $\phi$-neighborhood,
    $F\subseteq B$, and we know $B\subseteq G'$. Then, since $E$ is
    monotone, $x\in \iexpr{E}{G'}(v)$. Next, by definition of
    $\phi$-neighborhood, $(v,p,x)\in B$. Since $B\subseteq G'$, also
    $x\in\iexpr{p}{G'}(v)$ as desired.
    
  \item $\phi$ is $\neg \lessthaneq(E,p)$. Analogous to the case where
    $\phi$ is $\neg \lessthan(E,p)$.
    
  \item $\phi$ is $\neg \uniquelang(E)$. Since $G,v\models\phi$, there
    exists $x_1\neq x_2\in\iexpr{E}{G}(v)$ such that $x_1\eqlang x_2$. As in
    the previous case, we must show that $x_1$ and $x_2$ are in
    $\iexpr{E}{G'}(v)$. By Proposition~\ref{prop:graphpaths}, for both
    $i=1,2$, we have $x_i\in \iexpr{E}{F_i}(v)$ with
    $F_i=\graph(\paths(E,G,v,x_i))$. By definition of
    $\phi$-neighborhood $F_i\subseteq B\subseteq G'$. Therefore
    $x_i\in\iexpr{E}{G'}(v)$ as desired.
    
  \item $\phi$ is $\neg \closed(P)$. As $G,v\models\phi$, there exists
    a property $p\not\in P$ and a node $x$ such that $(v,p,x) \in
    G$. By definition $(v,p,x) \in B(v,G,\phi) \subseteq G'$. Hence,
    $G',v\models \phi$ as desired.
  \end{itemize}
\end{proof}

\section{Shape fragments in SPARQL}

\subsection{Proof of Lemma~\ref{lemE}} \label{appE}

Proceeding by induction on the structure of $E$, we describe
$Q_E$ in each case.

\lstset{
  basicstyle=\ttfamily\small,
  mathescape
}

\begin{itemize}
\item $E$ is a property name $p$.
\begin{lstlisting}
    SELECT (?s AS ?t) ?s ($p$ AS ?p) ?o (?o AS ?h)
    WHERE { ?s $p$ ?o. }
\end{lstlisting}
\item $E$ is $E_1?$.
\begin{lstlisting}
    SELECT ?t ?s ?p ?o ?h
    WHERE {
      { $Q_{E_1}$ }
      UNION 
      {
        SELECT (?v AS ?t) (?v AS ?h)
        WHERE { { ?v ?_p1 ?_o1 } UNION { ?_s2 ?_p2 ?v } }
      }
    }
\end{lstlisting}
\item $E$ is $E_1^-$.
\begin{lstlisting}
    SELECT (?h AS ?t) ?s ?p ?o (?t as ?h)
    WHERE {
      $Q_{E_1}$
    }
\end{lstlisting}
\item $E$ is $E_1 \cup E_2$.
\begin{lstlisting}
    SELECT ?t ?s ?p ?o ?h 
    WHERE {
      { $Q_{E_1}$ }
      UNION
      { $Q_{E_2}$ }
    }
\end{lstlisting}
\item $E$ is $E_1 \comp E_2$. 
\begin{lstlisting}
    SELECT ?t ?s ?p ?o ?h 
    WHERE {
      {
        {
          SELECT ?t ?s ?p ?o (?h AS ?h1)
          WHERE { $Q_{E_1}$ }
        }.
        {
          SELECT (?t AS ?h1) ?h
          WHERE { ?t $E_2$ ?h }
        } 
      }
      UNION
      {
        {
          SELECT ?t (?h AS ?h1)
          WHERE { ?t $E_1$ ?h }
        }.
        {
          SELECT (?t AS ?h1) ?s ?p ?o ?h
          WHERE { $Q_{E_2}$ }
        }
      }
    }
\end{lstlisting}
\item $E$ is $E_1^*$.
\begin{lstlisting}
    SELECT ?t ?s ?p ?o ?h
    WHERE {
      {
        ?t $E_1^*$ ?x1.
        ?x2 $E_1^*$ ?h.
        {
          SELECT (?t AS ?x1) ?s ?p ?o (?h AS ?x2)
          WHERE { $Q_{E_1}$ }
        }
      }
      UNION
      { 
        SELECT (?v AS ?h) (?v AS ?t)
        WHERE { { ?v ?_p1 ?_o1 } UNION { ?_s2 ?_p2 ?v } }
      }
    }
\end{lstlisting}
\end{itemize}

\subsection{Proof of Lemma~\ref{lemB}} \label{appB}

As always we work in the context of a schema $H$.
We assume $\phi$ is put in negation normal form and proceed by
induction as in the proof of the Sufficiency Lemma.

Note that $Q_\phi$ should not merely check conformance of nodes
to shapes, but actually must return the neighborhoods.  Indeed,
that conformance checking in itself is possible in SPARQL (for
nonrecursive shapes) is well known; it was even considered
for recursive shapes \cite{shaclsparql}.  Hence, in the
constructions below, we use an auxiliary SPARQL query
$\mathit{CQ}_\phi(?v)$ (C for conformance)
which returns, on every RDF graph $G$, the
set of nodes $v \in N(G)$ such that $G,v \models \phi$.  

We now describe $Q_\phi$ for all the cases in the following.

\begin{itemize}
\item $\phi$ is $\top$: empty
\item $\phi$ is $\hasvalue(c)$: empty
\item $\phi$ is $\test(t)$: empty
\item $\phi$ is $\closed(P)$: empty
\item $\phi$ is $\hasshape(s)$: $Q_{\sdef sH}$
  
\item $\phi$ is $\phi_1 \land \phi_2$ or $\phi_1 \lor \phi_2$:
\begin{lstlisting}
    SELECT ?v ?s ?p ?o
    WHERE {
      { $\mathit{CQ}_\phi$ } .
      { $Q_{\phi_1}$ }
      UNION
      { $Q_{\phi_2}$ }
    }
\end{lstlisting}
  
\item $\phi$ is $\geqn{n}{E}{\phi_1}$:
\begin{lstlisting}
    SELECT (?t AS ?v) ?s ?p ?o
    WHERE {
      { 
        { SELECT (?v AS ?t) WHERE { $\mathit{CQ}_\phi$ } } .
        { $Q_E$ } . 
        { SELECT (?v AS ?h) WHERE { $\mathit{CQ}_{\phi_1}$ } }
      } UNION
      {
        { SELECT (?v AS ?t) WHERE { $\mathit{CQ}_\phi$ } } .
        ?t $E$ ?h .
        {
          SELECT (?v AS ?h) ?s ?p ?o
          WHERE { { $Q_{\phi_1}$ } . { $\mathit{CQ}_{\phi_1}$ }} 
        }
      }. 
    }
\end{lstlisting}

\item $\phi$ is $\leqn{n}{E}{\phi_1}$:
\begin{lstlisting}
    SELECT (?t AS ?v) ?s ?p ?o
    WHERE {
      { 
        { SELECT (?v AS ?t) WHERE { $\mathit{CQ}_\phi$ } } .
        { $Q_E$ } . 
        { SELECT (?v AS ?h) WHERE { $\mathit{CQ}_{\neg \phi_1}$ } }
      } UNION
      { 
        { SELECT (?v AS ?t) WHERE { $\mathit{CQ}_\phi$ } } .
        ?t $E$ ?h .
        {
          SELECT (?v AS ?h) ?s ?p ?o
          WHERE { { $Q_{\neg \phi_1}$ } . { $\mathit{CQ}_{\neg \phi_1}$ }} 
        }
      }
    }
\end{lstlisting}
  
\item $\phi$ is $\forall E.\phi_1$:
\begin{lstlisting}
    SELECT (?t AS ?v) ?s ?p ?o
    WHERE {
      { 
        { SELECT (?v AS ?t) WHERE { $\mathit{CQ}_\phi$ } } .
        { $Q_E$ }  
      } UNION
      { 
        { SELECT (?v AS ?t) WHERE { $\mathit{CQ}_\phi$ } } .
        ?t $E$ ?h .
        {
          SELECT (?v AS ?h) ?s ?p ?o
          WHERE { $Q_{\phi_1}$ } 
        }
      }
    }
\end{lstlisting}
  
\item $\phi$ is $\eq(E,p)$:
\begin{lstlisting}
    SELECT (?t AS ?v) ?s ?p ?o 
    WHERE {
      { SELECT (?v AS ?t) WHERE { $\mathit{CQ}_\phi$ } } . 
      {
        { $Q_E$ } UNION { $Q_{p}$ }
      } 
    }
\end{lstlisting}

\item $\phi$ is $\eq(\id,p)$:
\begin{lstlisting}
    SELECT ?v (?s AS ?v) ($p$ AS ?p) (?v AS ?o) 
    WHERE {
      { $\mathit{CQ}_\phi$ } .
      ?v $p$ ?v
    }
\end{lstlisting}

\item $\phi$ is $\disj(E,p)$: empty

\item $\phi$ is $\disj(\id,p)$: empty
  
\item $\phi$ is $\lessthan(E,p)$: empty

\item $\phi$ is $\lessthaneq(E,p)$: empty

\item $\phi$ is $\uniquelang(E)$: empty
  
\item $\phi$ is $\neg \top$: empty

\item $\phi$ is $\neg \hasvalue(c)$: empty
  
\item $\phi$ is $\neg \test(t)$: empty

\item $\phi$ is $\neg \hasshape(s)$: $Q_{\neg \sdef sH}$

\item $\phi$ is $\neg \closed(P))$:
\begin{lstlisting}
    SELECT ?v (?v AS ?s) ?p ?o
    WHERE {
      { $\mathit{CQ}_\phi$ } .
      ?v ?p ?o.
      FILTER (?p NOT IN $P$)
    }
\end{lstlisting}

\item $\phi$ is $\neg \eq(E,p)$:
\begin{lstlisting}
    SELECT (?t AS ?v) ?s ?p ?o
    WHERE {
      { SELECT (?v AS ?t) WHERE { $\mathit{CQ}_\phi$ } } .
      {
        { { $Q_{E}$ } MINUS { ?t $p$ ?h } }
        UNION
        { { $Q_{p}$ } MINUS { ?t $E$ ?h } }
      }
    }
  \end{lstlisting}

\item $\phi$ is $\neg \eq(\id,p)$:
\begin{lstlisting}
    SELECT ?v (?v AS ?s) ($p$ AS ?p) (?v AS ?o)
    WHERE {
      { $\mathit{CQ}_\phi$ } .
      { ?v $p$ ?o }
      FILTER (?o != ?v)
    }
\end{lstlisting}

\item $\phi$ is $\neg \disj(E,p)$:
\begin{lstlisting}
    SELECT (?t AS ?v) ?s ?p ?o
    WHERE {
      { SELECT (?v AS ?t) WHERE { $\mathit{CQ}_\phi$ } } .
      {
        { { $Q_{E}$ } . { ?t $p$ ?h } }
        UNION
        { { $Q_{p}$ } . { ?t $E$ ?h } }
      }
    }
\end{lstlisting}

\item $\phi$ is $\neg \disj(\id,p)$:
\begin{lstlisting}
  SELECT ?v (?v AS ?s) ($p$ AS ?p) (?v AS ?o)
  WHERE {
    { $\mathit{CQ}_\phi$ } .
    ?v $p$ ?v
  }
\end{lstlisting}

\item $\phi$ is $\neg \lessthan(E,p)$:
\begin{lstlisting}
    SELECT (?t AS ?v) ?s ?p ?o
    WHERE {
      { SELECT (?v AS ?t) WHERE { $\mathit{CQ}_\phi$ } } .
      {
        { { $Q_{E}$ } . { ?t $p$ ?h2 } FILTER (! ?h < ?h2) }
        UNION
        { { $Q_{p}$ } . { ?t $E$ ?h2 } FILTER (! ?h2 < ?h) }
      }
    }
\end{lstlisting}

\item $\phi$ is $\neg \lessthaneq(E,p)$:
\begin{lstlisting}
    SELECT (?t AS ?v) ?s ?p ?o
    WHERE {
      { SELECT (?v AS ?t) WHERE { $\mathit{CQ}_\phi$ } } .
      {
        { { $Q_{E}$ } . { ?t $p$ ?h2 } FILTER (! ?h <= ?h2) }
        UNION
        { { $Q_{p}$ } . { ?t $E$ ?h2 } FILTER (! ?h2 <= ?h) }
      }
    }
\end{lstlisting}

\item $\phi$ is $\neg \uniquelang(E)$:
\begin{lstlisting}
    SELECT (?t AS ?v) ?s ?p ?o
    WHERE {
      { SELECT (?v AS ?t) WHERE { $\mathit{CQ}_\phi$ } } .
      { $Q_{E}$ } . { ?t $E$ ?h2 }
      FILTER (?h != ?h2 && lang(?h) = lang(?h2))
    }
\end{lstlisting}
\end{itemize}

\section{Proof of Proposition~\ref{proptpfexp}}

That the seven forms of TPF mentioned in the proposition can be
expressed as shape fragments was already shown in the main body
of the paper.  So it remains to show that all other forms of TPF
are not expressible as shape fragments.  Since, for any finite set
$S$ of shapes, we can form the disjunction $\bigvee S$ of all
shapes in $S$, and $\frag(G,S)=\frag(G,\{\bigvee S\})$ for any
graph $G$, it suffices to consider single shapes $\phi$ instead of
finite sets of shapes.  We abbreviate $\frag(G,\{\phi\})$ to
$\frag(G,\phi)$.

Formally, let $Q=(u,v,w)$ be a triple pattern, i.e., $u$, $v$ and
$w$ are variables or elements of $N$. Let $V$ be the set of
variables from $\{u,v,w\}$ to $N$.  A solution mapping is a
function $\mu : V \to N$.  For any node $a$, we agree that
$\mu(a)=a$.  Then the TPF query
$Q$ maps any input graph $G$ to the subset $$ Q(G) =
\{(\mu(u),\mu(v),\mu(w)) \mid \mu : V \to N \enne
(\mu(u),\mu(v),\mu(w)) \in G\}. $$
We now say that a shape $\phi$ \emph{expresses} a TPF query $Q$
if $\frag(G,\phi)=Q(G)$ for every graph $G$.

We begin by showing:

\begin{lemma} \label{lempnm}
  Let $G$ be an RDF graph and let $\phi$ be a shape.  Assume
  $\frag(G,\phi)$ contains a triple $(s,p,o)$ where $p$ is not
  mentioned in $\phi$.  Then $\frag(G,\phi)$ contains all triples
  in $G$ of the form $(s,p',o')$, where $p'$ is not mentioned in
  $\phi$.
\end{lemma}
\begin{proof}
  Since shape fragments are unions of neighborhoods, it suffices
  to verify the statement for an arbitrary neighborhood
  $B(v,G,\phi)$.  This is done
  by induction on the structure of the negation normal form of
  $\phi$.  In almost all cases of
  Table~\ref{tabdefrag}, triples from $B(v,G,\phi)$ come from
  $E$-paths, with $E$ mentioned in $\phi$; from $B(v,G,\psi)$,
  with $\psi$ a subshape of $\phi$ or the negation thereof; or
  involve a property $p$ clearly mentioned in $\phi$.
  Triples of the first kind never have a property not mentioned
  in $\phi$, and triples of the second kind satisfy the statement
  by induction.

  The only remaining case is $\neg \closed(P)$. Assume $(v,p,x)$
  is in the neighborhood, and let $(v,p',x') \in G$ be a triple
  such that $p'$ is not mentioned in $\phi$.  Then certainly
  $p'\notin P$, so $(v,p',x')$ also belongs to the neighborhoods,
  as desired.
\end{proof}

Using the above Lemma, we give:

\begin{proof}[Proof of Proposition~\ref{proptpfexp}]
  Consider the TPF $Q=(?x,?x,?y)$ and assume there exists a shape
  $\phi$ such that $Q(G)=\frag(G,\phi)$ for all $G$.  Consider
  $G=\{(a,a,b),(a,c,b)\}$, where $a$ and $c$ are not mentioned in
  $\phi$.  We have $(a,a,b) \in Q(G)$ so $(a,a,b) \in
  \frag(G,\phi)$. Then by Lemma~\ref{lempnm}, also $(a,c,b) \in
  \frag(G,\phi)$.  However, $(a,c,b) \notin P(G)$, so we arrive
  at a contradiction, and $\phi$ cannot exist.

Similar reasoning can be used for  
all other forms of TPF not covered by the
  proposition.  Below we give the table of these TPFs $Q$, where
  $c$ and $d$ are arbitrary IRIs, possibly equal, and $?x$ and
  $?y$ are distinct variables.  The right column lists the
  counterexample graph $G$ showing that $Q(G) \neq \frag(G,\phi)$.
  Importantly,
  the property ($a$ or $b$) of the triples in $G$ is always chosen so
  that it is not mentioned in $\phi$, and moreover, $a$, $b$ and
  $e$ are distinct and also distinct from $c$ and $d$.
  $$ \begin{array}{ll}
Q & G \\
\hline
    (?x,?y,?x) & \{(a,b,a),(a,b,c)\} \\
    (?x,?y,?y) & \{(a,b,b),(a,b,c)\} \\
    (?x,?x,?x) & \{(a,a,a),(a,a,b)\} \\
    (?x,?y,c) & \{(a,b,c),(a,b,d)\} \\
    (?x,?x,c) & \{(a,a,c),(a,a,d)\} \\
    (?x,?y,?y) & \{(a,b,b),(a,b,c)\} \\
    (c,?x,?x) & \{(c,a,a),(c,a,b)\} \\
    (c,?x,d) & \{(c,a,d),(c,a,e)\}
  \end{array} $$
\end{proof}

\end{document}